\documentclass[onecolumn]{article}
\usepackage{graphicx}
\usepackage{amsmath}
\usepackage{epsfig}
\usepackage{psfig}
\usepackage{amsfonts}
\usepackage{amssymb}

\begin{document}

\title{Entropy Crisis, Ideal Glass Transition and Polymer Melting: Exact Solution on
a Husimi Cactus}
\author{Andrea Corsi and P. D. Gujrati\\Department of Polymer Science and Department of Physics\\The University of Akron, Akron, OH, 44325}
\date{\today}
\maketitle
\begin{abstract}
We introduce an extension of the lattice model of melting of semiflexible
polymers originally proposed by Flory. Along with a bending penalty
$\varepsilon$, present in the original model and involving three sites of the
lattice, we introduce an interaction energy $\varepsilon_{\mathrm{p}}$,
corresponding to the presence of a pair of parallel bonds and an interaction
energy $\varepsilon_{\mathrm{h}}$, associated with a hairpin turn. Both these
new terms represent four-site interactions. The model is solved exactly on a
Husimi cactus, which approximates a square lattice. We study the phase diagram
of the system as a function of the energies. For a proper choice of the
interaction energies, the model exhibits a first-order melting transition
between a liquid and a crystalline phase at a temperature $T_{\text{M}}$. The
continuation of the liquid phase below $T_{\text{M }}$gives rise to a
supercooled liquid, which turns continuously into a new low-temperature phase,
called metastable liquid, at $T_{\text{MC}}<T_{M}$. This liquid-liquid
transition seems to have some features that are characteristic of the critical
transition predicted by the mode-coupling theory. The exact calculation
provides a thermodynamic justification for the entropy crisis (entropy
becoming negative), generally known as the Kauzmann paradox, caused by the
rapid drop of the entropy near the Kauzmann temperature. It occurs not in the
supercooled liquid, but in the metastable liquid phase since its Helmholtz
free energy equals the absolute zero equilibrium free energy at a positive
temperature. A continuous ideal glass transition occurs to avoid the crisis
when the metastable liquid entropy, and not the excess entropy, goes to zero.
The melting transition in the original Flory model, corresponding to the
vanishing of the four-site interactions, appears as a tricritical point of the
new model.
\end{abstract}

\section{INTRODUCTION}

Flory's model of crystallization of semiflexible linear polymers [1,2] is well
known and describes a completely inactive crystal (CR) at low temperatures [3]
within the approximations developed independently by Flory [2], and by Huggins
[4]. There is a first-order melting to a disordered equilibrium liquid (EL)
phase, which has made it an attractive model to justify the Kauzmann paradox
[5] and the ideal glass transition [6] in supercooled liquids (SCL) [5-31].
The ideal glass transition in fragile supercooled liquids [7,9,17] is believed
to be a manifestation of a \textit{rapid drop} in the (configurational)
entropy [32] near the Kauzmann temperature \textit{T}$_{\mathrm{K}}$, and has
been a topic of many recent investigations [24-31]. The entropy in this work
will always refer to the configurational entropy [32]. There are competing
theories, both for and against an ideal glass transition, and the situation is
far from clear [7]. Even the nature of the melting transition in the Flory
model is in dispute [10-13,27], mainly because the Gujrati-Goldstein
excitations [3,10,11] destroy the complete inactive nature of the crystal
phase. The present work is motivated by this confused state of the field, and
provides a convincing argument in favor of an ideal glass transition at a
finite non-zero temperature. In order to substantiate our claims, we need to
consider an \textit{extension} of the original Flory model of melting. We also
clarify the nature of the melting transition in the Flory model [10,11,13,27].
Our conclusions are based on exact calculations. Some of the preliminary
results have appeared earlier [28]. The present work provides the missing
details in Ref. 28.

According to the paradox\textit{,} originally introduced by Kauzmann [5], the
extrapolated entropy \textit{S}$_{\mathrm{SCL}}$(\textit{T}) of the
supercooled liquid becomes less than the entropy \textit{S}$_{\mathrm{CR}}%
$(\textit{T}) of the corresponding CR at temperatures below the glass
transition temperature \textit{T}$_{\mathrm{G}}$. It is a common belief
[5-7,14,15] that the entropy of SCL \textit{cannot} be less than the entropy
of CR. However, it is worth noting [7,28,31] that there is no thermodynamic
requirement for or against this. It also does not violate the third law of
thermodynamics. However, treating this possibility as a paradox, now
conventionally known as the \textit{Kauzmann} \textit{paradox} or
\textit{catastrophe}, Kauzmann suggested that the system would either
crystallize spontaneously [5] or undergo an ideal glass transition [5-10,17]
to avoid the paradox.

The existence of a glass transition caused by the above paradox has been
originally justified [8] only in the Flory model of melting applied to linear
polymers that are long. The approximate calculation [8] shows that the
CR\ phase is completely inactive (zero specific heat, and zero entropy). The
supercooled liquid avoids the Kauzmann catastrophe by undergoing a continuous
transition called the ideal glass transition. This pivotal work enshrined the
Kauzmann catastrophe as probably the most important mechanism behind the glass transition.

It should be stressed that the glass transition is ubiquitous and is also seen
in small molecules. \textit{However, no model calculation exists that
demonstrates the paradox for small molecules. } Unfortunately, the
approximations used by Gibbs and DiMarzio [8] have subsequently been
rigorously proven to be incorrect, and unreliable [10-13], casting doubts on
their primary conclusion of the existence of the Kauzmann paradox. Thus, there
is currently no justification for the paradox as the root cause for the ideal
glass transition, at least in long polymers.

The current work is motivated by a desire to see if we can, nevertheless,
justify a thermodynamic basis of the ideal glass transition in very long
linear polymers. To this end, we perform an exact calculation. We should point
out that recently we have discovered the existence of an ideal glass
transition in a model of simple fluids [31(a)] and of dimers [31(b)]. However,
this work deals only with long polymers. As discussed elsewhere [28, 31], the
ideal glass transition in our view comes about not due to the originally
proposed Kauzmann paradox caused by \textit{S}$_{\mathrm{SCL}}$(\textit{T})
$<$%
\textit{S}$_{\mathrm{CR}}$(\textit{T}), but because of the \textit{entropy
crisis }when the entropy of the state becomes \textit{negative}. A negative
entropy implies that there \textit{cannot} be any realizable configuration of
the system, which is impossible as there must be at least one configuration
for the system to exist in Nature. Thus, in the following, we interpret the
Kauzmann paradox not in the original sense, but in the sense of the above
entropy crisis.

In the Flory model, a polymer chain is assumed to consist of $n\mathit{\ }%
$equal segments, each with the same size as the solvent molecule. Each site of
the lattice is occupied by either a polymer segment or a solvent molecule, and
the excluded volume effects are taken into account by requiring a site to be
occupied only once, either by a solvent molecule or a polymer segment. We can
also think of the solvent as representing voids in the system. The polymer
chain occupies a contiguous sequence of all the lattice sites connected by
polymer bonds. For concreteness and ease of discussion, we take the lattice to
be a square lattice, which approximates a tetrahedral lattice on which the
model is supposed to be defined. Both lattices have the same coordination
number $q$=4. At every site, the polymer chain can assume either a trans
conformation (the conformation is related to the state of two consecutive
bonds), when the consecutive bonds are collinear, or one of the two possible
gauche conformations, when the polymer chain bends. For a semiflexible polymer
chain, every gauche conformation has an energy penalty $\varepsilon$\thinspace
compared to a trans conformation. We set the energy for a trans conformation
to be zero. The total energy of the system in a configuration on a lattice of
$N$ sites is
\begin{equation}
\mathcal{E=}\mathit{N}_{\mathrm{g}}\varepsilon, \tag*{(1)}\label{eqn1}%
\end{equation}
where $\mathit{N}_{\mathrm{g}}$ is the number of gauche conformations present
in the configuration of the system. This interaction involves three
consecutive molecules of the chain and is the only one considered in the Flory model.

No interaction between non-consecutive portions of the same polymer chain or
between different polymer chains is taken into account in the Flory model
since, according to Flory [2], the crystallization of polymers is not due to
intermolecular interactions but due to internal ordering/disordering and
excluded volume interactions. Both the Flory [2] and the Huggins [4]
approximations predict that the (configurational) entropy \textit{S}%
(\textit{g}) of the polymer chain for a given fraction $g\equiv N_{\mathrm{g}%
}/N$ of gauche bonds goes to zero at a critical value $\ \mathit{g}_{0}$
(where $\mathit{g}_{0}$ is 0.45 in the Flory approximation [2,10(a)] and 0.227
in the Huggins approximation [4,10(b)]). Correspondingly, the predicted
entropy of the system is zero for any $g\leq g_{0}$ and gives rise to the
inactive phase for $g\leq g_{0}$. The result of the calculation is shown
schematically in Fig. 1. The system is in a disordered liquid phase EL at
temperatures higher than the melting temperature \textit{T}$_{\mathrm{M}}$
(curve BM). At \textit{T}$_{\mathrm{M}}$, the system undergoes a first order
transition to a completely inactive ordered CR, characterized by a zero free
energy and a zero density \textit{g} (portion MO). There is a discontinuity in
\textit{g} at \textit{T}$_{\mathrm{M}}$. The results due to Flory and Huggins
are qualitatively similar; the main difference is that Flory's estimate of
\textit{T}$_{\mathrm{M}}$ is about four times higher than that due to Huggins
[10(b)]. However, the simulations [12,27] strongly support the presence of the
Gujrati-Goldstein excitations that destroy the inactive crystal at low
temperatures, but the nature of the melting transition remains uncertain,
which makes the mathematical extrapolation MA representing the supercooled
liquid [8] questionable. In particular, it is not clear if the extrapolation
of the exact result would give a non-zero temperature where $\mathit{S}%
(\mathit{T})$ would go to zero but where $g>0$.%
\begin{figure}
[ptb]
\begin{center}
\includegraphics[width=5in]
{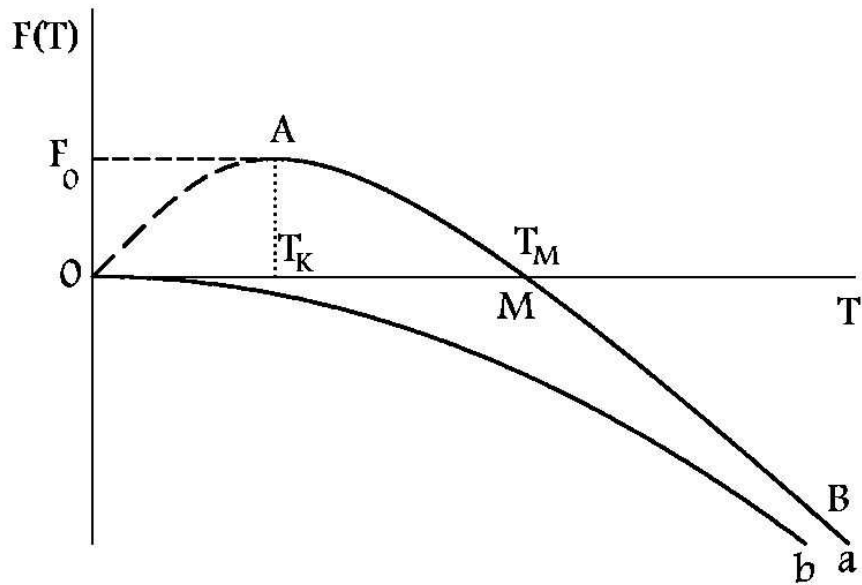}%
\caption{Free energy vs. temperature for a semiflexible polymer: (a) Flory's
calculation; (b) Gujrati-Goldstein upper bound.}%
\end{center}
\end{figure}

Rigorous lower bounds on $\mathit{S}(\mathit{g})$ per particle (and hence
upper bounds for the free energy) as a function of $g$ have been obtained
[10,11]. Gujrati and Goldstein were able to prove that the entropy per segment
of the chain in the case of a single polymer chain that occupies all the sites
of the lattice (the Hamilton walk limit) satisfies
\begin{equation}
\mathit{S}(g)\geq\left(  \frac{g}{8}\right)  \ln(\frac{4}{g}-3).
\tag*{(2)}\label{eqn2}%
\end{equation}
Hence, \textit{S} is positive for any value of \textit{g}
$>$%
0, as it surely must be, in contrast with the results obtained by Flory.
Bounds are also available for the case of finite-length polymers [11]. The
above bound (2) implies that the equilibrium free energy of the system is
never zero for $T>0$, see curve (b) in Fig. 1, and that the system is never
completely ordered at any finite temperature.

While the results due to Gujrati and Goldstein clearly show that the
approximations of Flory and Huggins do not give a satisfactory description of
the system, they just provide an upper bound for the equilibrium free energy
of the system; nothing is known about the correct equilibrium entropy.
Therefore, it is still unknown what the actual behavior of the free energy is,
which is needed to obtain the continuation of the SCL liquid phase. The bounds
do not say anything about the extrapolated (i.e. continued) SCL free energy or
entropy. The knowledge of the reliable free energy form is fundamental in
order to understand if there is a phase transition of any kind in the system
at any finite temperature. In particular, it is not clear if the model has a
first-order melting transition. It should be recalled that one can usually
continue the free energy only past a first-order transition, and not a second
order transition due to the singularity in the latter case. If there is no
melting transition with a latent heat, then there may be no SCL liquid below
the melting transition. In this case, there would be no validity to the
Gibbs-DiMarzio conjecture of a \ glass transition in the SCL liquid. Thus, an
exact calculation is highly desirable.

In recent years, the study of the glass transition has been stimulated by the
development of the mode-coupling theory (MCT) [20-22]. This theory was
developed in the first place for simple liquids but has been applied to
polymers also [20]. The MCT studies the evolution of the density
autocorrelation function that can be measured in scattering experiments or
calculated in computer simulations and is, therefore, of practical interest.
The main result of this theory is the prediction of a critical temperature
\textit{T}$_{\mathrm{MC}}$, above the glass transition temperature,
corresponding to a crossover in the dynamics of the system. At \textit{T}%
$_{\mathrm{MC}}$, the correlation time of the system (the segmental relaxation
time in the case of a polymeric glass) diverges with a power law just as one
observes near a critical point:
\begin{equation}
\tau\propto\left(  T-T_{\mathrm{MC}}\right)  ^{-\gamma}. \tag*{(3)}%
\label{eqn3}%
\end{equation}

Many neutron and light scattering experiments [20] have shown that the MCT is
able to predict at least qualitatively the spectra of low molecular weight
materials. Most of the systems for which MCT gives a good description of the
dynamics (at least qualitatively) belong to the class of fragile glass
formers. The theory has not been tested extensively with polymers that have
large molecular weight but at the same time have been shown to be the most
fragile systems yet identified [21-23].

Recent activities [24-26] have tried to export the progresses made in the
field of spin glasses [33] to the field of real glasses. Even though the
replica trick is clearly unphysical [34,35], this approach has been extended
to the case of real glasses. The replica approach has been applied to many
Lennard-Jones glasses and the results have been interesting [24-26]. They
provide some justification for the ideal glass transition. This may also be
the case for polymers, which is the focus of this study.

Despite the wide interest in the subject, there is still no comprehensive
understanding of the nature of the vitrifying SCL and its relationship with
CR, the mechanism responsible for the rapid entropy loss near \textit{T}%
$_{\mathrm{G}}$, and the nature of the ideal glass transition. It would also
be interesting to see if there is a possible thermodynamic basis for the
critical (and apparently a mode-coupling) transition in SCL's.

In order to obtain a thermodynamic justification for all these phenomena, we
consider in detail in this work a very simple limiting case. The solvent
density will be taken to be identically zero. Thus, we consider an
\textit{incompressible} pure system. The effect of free volume is treated in
separate publications [29-31]. We also consider the limiting case of a single
chain covering the entire lattice. Such a limit is conventionally known as the
Hamilton walk limit [10,11]. The case of many chains of finite lengths is
considered elsewhere [29-31]. To obtain a first order melting, we have to
extend the Flory model of melting, as described below. We have substituted the
original square lattice with a Husimi cactus (Fig. 2) on which the original
problem is solved \textit{exactly}. This is the only \textit{approximation} we
make. The results of this calculation for the case of a special interaction
have been reported earlier [28] but details were not given. The present work
also provides the missing details.%
\begin{figure}
[t]
\begin{center}
\includegraphics[width=5in]
{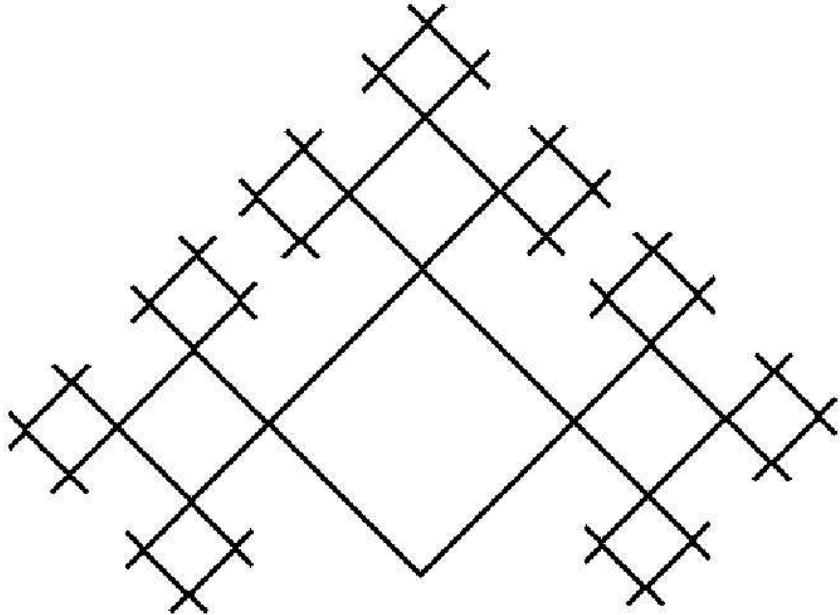}%
\caption{Upper half of a Husimi cactus of generation $m=3$. The dangling bonds
outside it show its connection through surface sites (not shown in the figure)
with the larger infinite cactus, as explained in the text.}%
\end{center}
\end{figure}

It has been previously shown [36] that the exact calculations on recursive
structures like the Husimi cactus are more reliable than conventional
mean-field calculations. In this approach, the problem is solved exactly,
taking into account \textit{all} correlations present on the recursive
lattice. In most cases, the real lattice is approximated by a tree structure.
Because of the tree nature, the correlations are weak. We have chosen the
Husimi cactus, obtained by joining two squares (Fig. 2) at each vertex, so
that the coordination number $q=4$. On a square lattice, there are also
squares that share a bond. Such squares are not present in the cactus. Thus,
the cactus should be thought of representing the \textit{checkerboard} version
of the square lattice, with the further provision that no closed loops of size
larger than the elementary square are present. The square cactus is chosen to
allow for the Gujrati-Goldstein excitations [10,11] that are important in
disordering the ideal crystal at absolute zero. \textit{The results from the
cactus calculation can be thought of as representing an approximate theory of
the model on a square lattice.}

The layout of the paper is as follows. In the next section, we introduce the
lattice model in terms of independent extensive quantities of interest. It is
the most general model provided we restrict these quantities to represent
pairs, triplets, and quadruplets of sites within each square. We also discuss
the general physics of the model. As said above, we use a square lattice for
simplicity to introduce the model, even though we eventually consider a Husimi
cactus, on which the calculations are exactly. In Sect. III, we explain the
recursive solution method on a Husimi cactus. We introduce 1-cycle and 2-cycle
solutions, representing the disordered and the ordered phase, respectively.
The results are presented in Sect. IV along with the discussion, and the final
section contains our conclusions.

\section{THE MODEL AND ITS PHYSICS}

\subsection{Independent extensive quantities}

We consider a square lattice of $N$ sites to focus our attention. We will
neglect surface effects. There are $N_{\text{B}}=$2$N$ lattice bonds, or
distinct pairs of sites. Let us describe the state of a square by the number
of polymer bonds $j$ in it. The bonds in the following refer to the polymer
bonds. Let $N_{\mathrm{S0}}$, and $N_{\mathrm{S1}}$ denote the number of
squares (S) with $j=0$, and 1, respectively; see Fig. 3.%
\begin{figure}
[tb]
\begin{center}
\includegraphics[width=5in]
{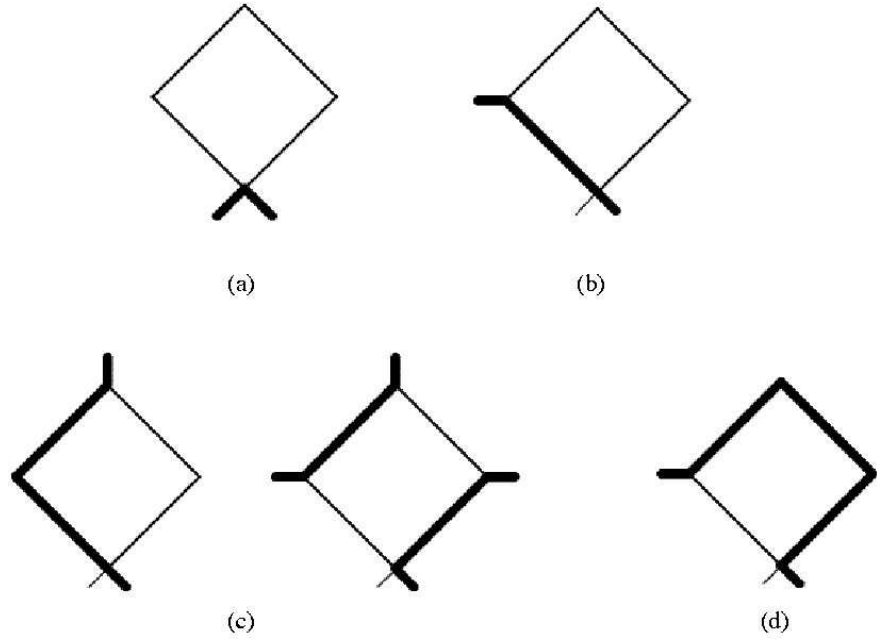}%
\caption{The possible states of a square in the lattice: (a) no bonds, (b) one
bond, (c) two bonds and (d) three bonds.}%
\end{center}
\end{figure}

For $j=2$, we distinguish between the case of parallel bonds (p), with the
number of such squares $N_{\mathrm{Sp}}$, and gauche (g) bonds, with the
number of such squares $N_{\mathrm{Sg}}$. The hairpin (h) turn corresponds to
$j=3$, with the number of such squares $N_{\mathrm{Sh}}$. No square is allowed
to have four bonds in it. Let $N_{\mathrm{t}}$ and $N_{\mathrm{g}}$ denote the
number of trans and gauche bonds, respectively, and $N_{\mathrm{p}}$ and
$N_{\mathrm{h}}$\ the number of pairs of parallel bonds and hairpin turns,
respectively, in a given configuration. We will also use them to represent
their average values, as there will be no confusion. It is easily seen that
the number of squares on a square lattice is $N_{\mathrm{S}}=N$. Let $B$
denote the number of polymer bonds, and $N_{\mathrm{mm}}$ the number of
unbonded monomer-monomer contacts. The following topological identities are
easily seen to hold:
\begin{equation}
N_{\text{\textrm{S}}}=N_{\mathrm{S0}}+N_{\mathrm{S1}}+N_{\mathrm{Sp}%
}+N_{\mathrm{Sg}}+N_{\mathrm{Sh}}, \tag*{(4)}%
\end{equation}%
\begin{equation}
2N_{\text{\textrm{mm}}}=4N_{\mathrm{S0}}+3N_{\mathrm{S1}}+2N_{\mathrm{Sp}%
}+2N_{\mathrm{Sg}}+N_{\mathrm{Sh}}, \tag*{(5)}\label{eqn 5}%
\end{equation}%
\begin{equation}
2B=N_{\mathrm{S1}}+2N_{\mathrm{Sp}}+2N_{\mathrm{Sg}}+3N_{\mathrm{Sh}},
\tag*{(6)}\label{eqn 6}%
\end{equation}%
\begin{equation}
N=N_{\mathrm{t}}+N_{\mathrm{g}},\text{ \ }2N=B+N_{\mathrm{mm}}, \tag*{(7)}%
\label{eqn 7}%
\end{equation}%
\begin{equation}
N_{\text{\textrm{p}}}=N_{\mathrm{Sp}}+N_{\mathrm{Sh}},\text{ \ }N_{\mathrm{g}%
}=N_{\mathrm{Sg}}+2N_{\mathrm{Sh}}. \tag*{(8)}\label{eqn 8}%
\end{equation}

As said earlier, the cactus represents the \textit{checkerboard} version of
the square lattice, so that the number of squares $N_{\text{S}}$ on both
lattices with the same number of sites $N$ are not the same; see Sect. III
also. For a square lattice, $N_{\text{S}}$= $N$; for the cactus, $N_{\text{S}%
}$ = $N/2$. However, the number of lattice bonds $N_{\text{B}}$ on both
lattices are the same. Because of this, Eq. (6) must be modified for the
cactus. Since each bond belongs to only one square in the cactus, we have:%

\begin{equation}
B=N_{\mathrm{S1}}+2N_{\mathrm{Sp}}+2N_{\mathrm{Sg}}+3N_{\mathrm{Sh}}.
\tag*{(6a)}%
\end{equation}
All other identities remain valid on the cactus.

\subsection{General Model}

Among the eleven extensive quantities, $N$, $B$, $N_{\mathrm{g}}$,
$N_{\mathrm{t}}$, $N_{\mathrm{mm}}$, $N_{\mathrm{S0}}$, $N_{\mathrm{S1}}$,
$N_{\mathrm{Sp}}$, $N_{\mathrm{Sg}}$, $N_{\mathrm{St}}$ and $N_{\mathrm{Sh}}$,
there are six independent geometrical relations; the second one in (7) is not
independent. In addition, for the Hamilton walk problem, we have $B=N$. Thus,
there are only four independent extensive quantities, which we take to be $N$,
$N_{\mathrm{g}}$, $N_{\mathrm{p}}$ and $N_{\mathrm{h}}$. One of these, the
lattice size $N$, will be used to define the partition function. The remaining
three independent quantities $N_{\mathrm{p}}$, $N_{\mathrm{g}}$ and
$N_{\mathrm{h}}$\ will be then used to define the configuration uniquely.
Corresponding to each of the quantities $N_{\mathrm{g}}$, $N_{\mathrm{p}}$ and
$N_{\mathrm{h}}$, there is an independent activity $w$, $w_{\mathrm{p}}$ and
$w_{\mathrm{h}}$, respectively, that will determine the partition function for
the Hamilton walk problem as
\begin{equation}
Z_{N}=\sum w^{N_{\mathrm{g}}}w_{\mathrm{p}}^{N_{\mathrm{p}}}w_{\mathrm{h}%
}^{N_{\mathrm{h}}}, \tag*{(9a)}\label{eqn9}%
\end{equation}
where the sum is over \textit{distinct configurations} obtained by all
possible values of $N_{\mathrm{g}}$, $N_{\mathrm{p}}$ and $N_{\mathrm{h}}%
$\ consistent with a fixed lattice size $N$. The activities $w$,
$w_{\mathrm{p}} $ and $w_{\mathrm{h}}$\ are determined by the three-site
bending penalty $\varepsilon$\ introduced by Flory in his model, an energy of
interaction $\varepsilon_{\mathrm{p}}$ associated with each parallel pair of
neighboring bonds, and an energy $\varepsilon_{\mathrm{h}}$\ for each hairpin
turn within each square as follows:%

\[
w=\exp\left(  -\beta\varepsilon\right)  ,\ w_{\mathrm{p}}=\exp\left(
-\beta\varepsilon_{\mathrm{p}}\right)  ,\ w_{\mathrm{h}}=\exp(-\beta
\varepsilon_{\mathrm{h}}).
\]
Here, $\beta$ is the inverse temperature \textit{T} in the units of the
Boltzmann constant. The original Flory model is obtained when the last two
interactions are absent. It should be stressed that $\varepsilon_{\mathrm{h}}$
is the excess energy associated to the configuration, once the energy of the
two bends and the pair of parallel bonds have been subtracted out. Both
$\varepsilon_{\mathrm{p}}$ and $\varepsilon_{\mathrm{h}}$\ are associated with
four-site interactions, since it is necessary to determine the state of four
adjacent sites to determine if a pair of parallel bonds or a hairpin turn is present.

The model is easily generalized to include free volume by introducing voids,
each of which occupies a site of the lattice. The number of voids $N_{0}$ is
controlled by the void activity $\eta$. The number $P$ of polymers is
controlled by another activity given by $H^{2}.$ The interaction between
nearest neighbor pairs $N_{\text{c}}$ of voids and the monomers of the
polymers determines the Boltzmann weight $w_{\text{c}}$. The partition
function of the extended model is given by%

\begin{equation}
Z_{N}=\sum\eta^{N_{0}}H^{2P}w_{\text{c}}^{N_{\text{c}}}w^{N_{\mathrm{g}}%
}w_{\mathrm{p}}^{N_{\mathrm{p}}}w_{\mathrm{h}}^{N_{\mathrm{h}}}, \tag*{(9b)}%
\end{equation}
where the sum is over distinct configurations consistent with the fixed
lattice of $N$ sites. Because the activity $H$ only determines the average
number of linear polymers, but not their individual sizes, the model in Eq.
(9b) describes polydisperse polymers [37], each of which must contain at least
one bond.

We now turn to our simplified model of the Hamilton walk ($P=1$, and
$\ N_{0}=0$). In this model, the energy of interaction in a given
configuration is given by
\begin{equation}
\mathcal{E}=\varepsilon N_{\mathrm{g}}+\varepsilon_{\mathrm{p}}N_{\mathrm{p}%
}+\varepsilon_{\mathrm{h}}N_{\mathrm{h}}=\varepsilon\left(  N_{\mathrm{g}%
}+aN_{\mathrm{p}}+bN_{\mathrm{h}}\right)  , \tag*{(10)}\label{eqn10}%
\end{equation}
where $a\equiv\varepsilon_{\mathrm{p}}/\varepsilon$, $b\equiv\varepsilon
_{\mathrm{h}}/\varepsilon$. The parameters can in principle assume positive
and negative values. However, we will restrict ourselves to $\varepsilon>0$ in
this work. The limit $\varepsilon=\varepsilon_{\text{p}}=\varepsilon
_{\text{h}}=0$ corresponds to a completely \textit{flexible} polymer problem,
which is of no interest to us here, as it corresponds to the infinite
temperature limit of our model. The limit, however, is of considerable
interest in the study of protein folding and has been investigated by several
workers [38]. In addition, we will focus mainly on the case $0<a<1$. A
positive $a$\ guarantees that parallel bond energy opposes the creation of
configurations in which pairs of parallel bonds are present and $a<1$ makes
the penalty for a pair of parallel bonds less than that for a gauche
conformation. This guarantees the presence of a crystalline phase at low
temperatures, as shown below.

\subsection{Ground State at $\mathit{T}$=0}

The physics of the model at absolute zero can be easily understood on general
grounds. We are interested in the thermodynamic limit $N\rightarrow\infty$. We
first consider $b=0$. For $a<1$, the ground state at $T=0$ has $N_{\mathrm{g}%
}=0$, $N_{\mathrm{p}}=N$, $N_{\mathrm{h}}=0$, as shown in Fig. 4(a). (The
labels R and L are related to the state of the sites as introduced in the next
section.) Thus, $\mathcal{E}=\varepsilon_{\text{p}}N$. This is what we will
call the perfect crystal at absolute zero. For $b\neq0$, the state in Fig.
4(a) remains the ground state as long as $2+b>0$. This condition ensures that
hairpin turns are not present. For $a>1$, and $b=0$, the ground state at $T=0$
has $N_{\mathrm{g}}=N$, $N_{\mathrm{p}}=0$, as shown in Fig. 4(b). Thus,
$\mathcal{E}=\varepsilon N$. This remains the ground state provided $a+b>0$,
which ensures that the hairpin turns are not present. Since our interest is to
have a crystal state as the equilibrium state at low temperatures, we would
only consider the earlier case $a<1$, with $2+b>0$.%
\begin{figure}
[tb]
\begin{center}
\includegraphics[width=5in]
{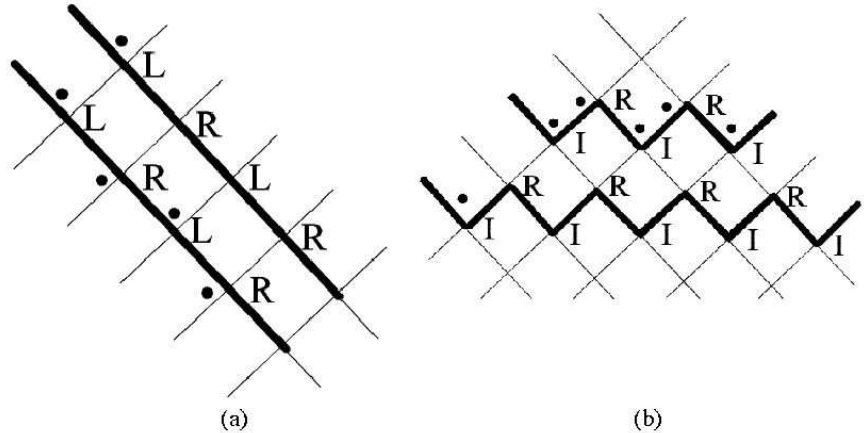}%
\caption{Possible configurations of the polymer chain at $T=0$: (a)
crystalline phase with $N_{\mathrm{g}}=0$ and $N_{\mathrm{p}}=N$; (b)
step-like configuration with $N_{\mathrm{p}}=0$ and $N_{\mathrm{g}}=N$. See
text for the explanation of the symbols.}%
\end{center}
\end{figure}

It should be recognized that the model considered here is defined on a
lattice. Thus, the ground state also possesses the symmetry of the lattice.
This symmetry is imposed by the lattice symmetry and is independent of the
model. Thus, this induced symmetry should not be confused with the point group
symmetry of a real crystal, which is brought about by the interactions in the
system. The symmetry in our model is due to the orientational order between
pairs of parallel bonds. It is because of the orientational order defining the
crystal in our model that we can obtain a continuous transition between the
crystal\ and the equilibrium liquid.

The lack of a point group symmetry of a real crystal in our model should not
be a taken as a serious limitation of the model, since our main goal is to
study the possibility of a glass transition in a supercooled liquid. The
determination of the supercooled liquid branch requires the continuation past
a first-order melting transition. Thus, the exact nature of the symmetry of
the CR phase is not as important as the existence of a discontinuous melting.

\section{RECURSIVE SOLUTION}

The Husimi cactus approximates the square lattice, as said above. Both have
the coordination number $q=4$, and the elemental squares as the smallest loop.
However, the most important reason for choosing the square cactus is that it
allows for hairpin turns that give rise to the Gujrati-Goldstein excitation in
the Flory model of melting. A Bethe lattice would be inappropriate for this
reason. As said above, the number of squares on the cactus is half of that on
a square lattice with the same number of sites $N$. This can also be easily
seen as follows by assuming homogeneity of the lattice. First, consider the
square lattice. Four squares meet at each site; however, each square will be
counted four times, due to its four corners, assuming homogeneity. Thus,
$N_{\mathrm{S}}=N$. On a cactus, only two squares meet, but each one is
counted four times as before. Hence, $N_{\mathrm{S}}=N/2$. Despite this,
$N_{\mathrm{B}}=2N$ on both lattices, only half of which are going to be taken
up by the Hamilton walk on both lattices.

A site is shared by four bonds and two squares $\Sigma$, and $\Sigma^{\prime}$
that are across from each other on the cactus. On the other hand, there are
two different pairs of such squares on a square lattice. In a formal sense, we
can imagine that each end of a bond contributes $%
\frac14
$\ of a site, and each corner of a square contributes $%
\frac12
$ of a site (on a square lattice each corner contributes $%
\frac14
$ of a site.). This formal picture will be useful in determining the nature of
a homogeneous cactus. To make the cactus homogeneous, we must consider it to
be part of a larger cactus. This is shown in Fig. 2, where we show a cactus of
generation $m=3$ with dangling bonds (each one ending with a surface site, not
shown in the figure) outside it to show its connection with the larger
infinite cactus. The latter has no boundary. A similar homogeneity hypothesis
associated with a Bethe lattice has been discussed in Ref. 37(b) to which we
refer the reader for further details. On a Bethe lattice, each dangling bond
was treated as a half-bond to ensure that $N_{\mathrm{B}}=qN/2$, where $q$ is
the coordination number of the Bethe lattice. For the case of the cactus, we
treat each pair of dangling bonds in Fig. 2 as a half-square, and each surface
site as a half-site to calculate the number of sites $N_{m}$, and the number
of squares $S_{m}$ for a cactus of $m $ generations. A trivial calculation
shows that
\begin{equation}
N_{m}=4\times3^{m}-2,\text{ \ \ }S_{m}=2\times3^{m}-1, \tag*{(11)}%
\label{eqn11}%
\end{equation}
so that $S_{m}=N_{m}/2$, as $m\rightarrow\infty$. Since each square
contributes 4 bonds, it is also evident that\ $N_{\mathrm{B}}=2N$ in the limit
of an infinite cactus. A detailed calculation of the quantities introduced
above is given in the Appendix.

Because of the above-mentioned homogeneity, a site is arbitrarily designated
as the origin of the cactus. Each square has one site, called the base site,
closer to the origin. The base site is given an index $m\geq0$ , the two sites
next to the base site within the square, called the intermediate sites, the
index $(m+1)$, and the remaining fourth site, called the peak site, the index
$(m+2)$. We will call this square an $m$th level square; it has its base at
the $m$th level and its peak at the $(m+2)$th level; see Fig. 5. The two lower
bonds in the $m$th square connected to the $m$th site are called the lower
bonds and the two upper bonds connected to the peak site are called the upper
bonds. The origin of the lattice is labeled as the $m=0$ level and the level
index $m$ increases as we move outwards from the origin. We can imagine
cutting the Husimi tree at an $m$th site into two parts, one of which does not
contain the origin if $m>0$. We call this the $m $th branch of the lattice and
denote it by $\mathcal{C}_{m}$. At the origin, we get two identical branches
each containing the origin. We will call each of those the $(m=0)$th branch.%
\begin{figure}
[tb]
\begin{center}
\includegraphics[width=3in]
{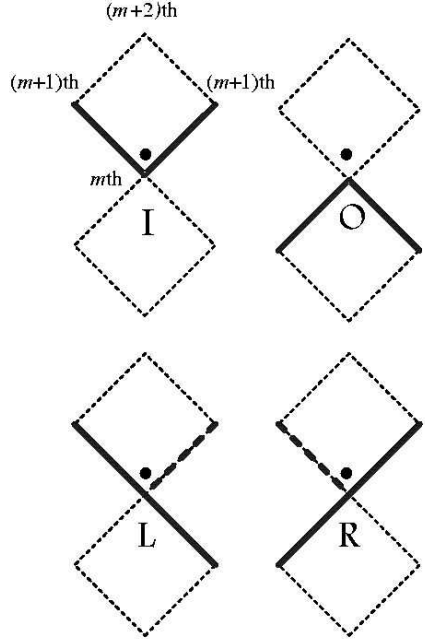}%
\caption{The four possible states of the polymer chain at any site at the
$m$th level of the lattice.}%
\end{center}
\end{figure}

We only consider parallel bonds and hairpin turns that are inside the squares,
since the cactus represents the checkerboard version of the square lattice.
Thus, each square can contribute only once to either $N_{\text{p}}$, or
$N_{\text{h}}.\;$This means that $N_{\text{p}}=N_{\text{S}}=N/2$ in the
perfect crystal at absolute zero. Similarly, it can be easily seen that the
maximum possible value of $N_{\text{h}}\;$is $2N_{\text{S}}/3=N/3$. \ \ \ \ \ \ \ \ \ 

\subsection{Recursion Relations}

We consider a linear polymer that covers all the sites of the Husimi cactus or
the square. Its configuration determines the state of the bonds in each
square. Consider a pair of two squares $\Sigma$, and $\Sigma^{\prime}\,$that
are across from each other. We distinguish $\Sigma$ by putting a filled dot
($\bullet$) just above the common site. We now face towards $\Sigma^{\prime}$
from within $\Sigma$ through this common site. The common site has been taken
as the base site in Fig. 5, but the following description is valid at any
site. The common site can assume 4 possible different states depending on the
state of the four bonds connected to it. Two of the bonds are in $\Sigma,$ and
the remaining two are in $\Sigma^{\prime}.$

\begin{enumerate}
\item  In the I state, both $\Sigma$- bonds are occupied by the polymer chain.
Since the polymer is linear, the two $\Sigma^{\prime}$-bonds must be
unoccupied by the polymer.

\item  In the O state, both $\Sigma$-bonds are unoccupied but both
$\Sigma^{\prime}$-bonds are occupied by the polymer chain.

\item  In the L state, only one of the $\Sigma$-bonds is occupied and the
polymer occupies the left bond in $\Sigma^{\prime}$ (we always think about
left and right as we face towards $\Sigma$').

\item  In the R state, only one of the $\Sigma$-bonds is occupied and the
polymer occupies the right bond in $\Sigma^{\prime}$.
\end{enumerate}

For the common site at $m=0$, the square $\Sigma^{\prime}$ in the above
classification is the square on the other side of the origin.

It is now easy to understand the labeling of the two configurations shown in
Fig. 4.

We are interested in the contribution of the portion of the $m$th branch
$\mathcal{C}_{m}$ of the lattice to the total partition function of the
system. This contribution is called the partial partition function (PPF) of
the branch. It is easy to see that the PPF depends on the state $\alpha$ of
the $m$th level site. We denote this PPF by $Z_{m}(\alpha)$. We now wish to
express $Z_{m}(\alpha)$\ recursively in terms of the PPF's of the two
intermediate sites and the peak site of the $m$th square. Following Gujrati
[36], the recursion relations can always be written in the form:
\begin{equation}
Z_{m}(\alpha)=\text{Tr}[\text{W}(\alpha,\left\{  \beta\right\}  )Z_{m+1}%
(\beta_{1})Z_{m+1}(\beta_{2})Z_{m+2}(\beta_{3})], \tag*{(12)}\label{eqn12}%
\end{equation}
where $\left\{  \beta\right\}  $ is the set of states $\beta_{i}$; the latter
states represent the possible states of the other three sites of the square,
and W$(\alpha,\left\{  \beta\right\}  )$ is the local Boltzmann weight of the
square due to conformation of the polymer chain inside the square.

Let us consider in detail the case in which the base site at the $m$th level
is in the I state. The three possible configurations that the polymer can
assume in the $m$th level square are shown in Fig. 6.%
\begin{figure}
[ptb]
\begin{center}
\includegraphics[width=5in]
{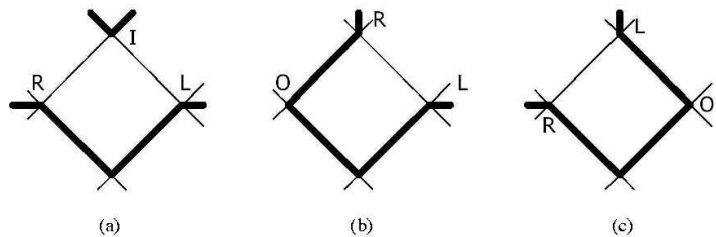}%
\caption{Possible configurations of the polymer chain when the $m$th level
site is in the I state.}%
\end{center}
\end{figure}
In this figure, L, R, I and O represent the possible state of the $(m+1)$th
and $(m+2)$th level sites and $w$ represents the weight of a bend. In order to
carefully account for statistical weights, a Boltzmann weight equal to $w$ is
considered only if the bend happens:

\begin{enumerate}
\item  At the $m$th level and at least one polymer bond at the level is inside
the square.

\item  At the $(m+1)$th or $(m+2)$th level, and both polymer bonds at the
level are inside the polymer.
\end{enumerate}

A weight $w_{\mathrm{p}}=w^{a}$ is considered for any configuration in which
two bonds are parallel to each other within the same square. We can,
furthermore, distinguish configurations in which two bonds are parallel to
each other from configurations, like those shown in Figs. 6(b)-(c), where
three consecutive bonds form a hairpin configuration. Whenever this
configuration is present, an additional weight $w_{\mathrm{h}}=w^{b}$ is introduced.

It is not important to know along which of the two lower bonds in the
$(m+1)$th or $(m+2)$th square does the polymer chain enter into the $m$th
square. In fact, even if the polymer undergoes a bend while moving from the
higher level square to the $m$th level square, the corresponding weight is
already taken into account into the partial partition function of the higher
level site.

It is important to consider always the state of a site moving through the
lattice towards the origin. In the configuration in Fig. 6(a), the
intermediate site on the left is in the R state since the polymer undergoes a
right turn after entering the square. The intermediate site on the right is in
the L state since the polymer undergoes a left turn after entering the square.
Finally, the peak site is in the I state because the polymer is coming from
the $(m+1)$th level square but not entering the $m$th level square. The
polymer undergoes one bend (at the base site) so that there is a weight $w$ to
be taken into account. Thus the contribution to the partition function coming
from this configuration is:
\begin{equation}
wZ_{m+1}(\mathrm{R})Z_{m+1}(\mathrm{L})Z_{m+2}(\mathrm{I}). \tag*{(13)}%
\label{eqn13}%
\end{equation}

In the configuration in Fig. 6(b), the intermediate site on the left is in the
O state since both the lower bonds in the corresponding $(m+1)$th square are
unoccupied. The intermediate site on the right is in the L state since the
polymer undergoes a left turn after entering the square. Finally, the peak
site is in the R state since the polymer undergoes a right turn after entering
the square. The polymer undergoes two bends (one at the base site and the
other one at the left intermediate site) so that there is a weight $w^{2}$ to
be taken into account. There is a pair of parallel bonds in the square and a
hairpin turn occurs so that a weight $w_{\mathrm{h}}w_{\mathrm{p}}$\ has also
to be taken into account. Thus, the contribution to the partition function
coming from this configuration is:
\begin{equation}
w^{2}w_{\mathrm{h}}w_{\mathrm{p}}Z_{m+1}(\mathrm{O})Z_{m+1}(\mathrm{L}%
)Z_{m+2}(\mathrm{R}). \tag*{(14)}\label{eqn14}%
\end{equation}

In the configuration in Fig. 6(c), the intermediate site on the right is in
the O state since both the lower bonds in this particular $(m+1)$th square are
unoccupied. The intermediate site on the left is in the R state since the
polymer undergoes a right turn after entering the square. Finally, the peak
site is in the L state since the polymer undergoes a left turn after entering
the square. The polymer undergoes two bends (one at the base site and the
other one at the right intermediate site) so that there is a weight $w^{2}$ to
be taken into account. We also have a pair of parallel bonds and a hairpin
turn to take into account in this case. Thus, the contribution to the
partition function coming from this configuration is:
\begin{equation}
w^{2}w_{\mathrm{h}}w_{\mathrm{p}}Z_{m+1}(\mathrm{O})Z_{m+1}(\mathrm{R}%
)Z_{m+2}(\mathrm{L}). \tag*{(15)}\label{eqn15}%
\end{equation}

The recursion relation for $Z_{m}(\mathrm{I})$, the partition function of the
$m $th branch of the Husimi tree given that the $m$th level site is in the I
state, is therefore given by:
\begin{align}
Z_{m}(\mathrm{I})  &  =w^{2}w_{\mathrm{h}}w_{\mathrm{p}}Z_{m+1}(\mathrm{O}%
)[Z_{m+1}(\mathrm{L})Z_{m+2}(\mathrm{R})+Z_{m+1}(\mathrm{R})Z_{m+2}%
(\mathrm{L})]\nonumber\\
&  +wZ_{m+1}(\mathrm{R})Z_{m+1}(\mathrm{L})Z_{m+2}(\mathrm{I}). \tag*{(16)}%
\end{align}

Considering the case in which the $m$th level site is in the O state, the
partial partition function $Z_{m}(\mathrm{O})$ for the O state can be written
as:
\begin{align}
Z_{m}(\mathrm{O})  &  =Z_{m+1}^{2}(\mathrm{I})Z_{m+2}(\mathrm{I}%
)+Z_{m+1}(\mathrm{I})\left[  Z_{m+1}(\mathrm{L})Z_{m+2}(\mathrm{R}%
)+Z_{m+1}(\mathrm{R})Z_{m+2}(\mathrm{L})\right] \nonumber\\
&  +wZ_{m+1}(\mathrm{R})Z_{m+1}(\mathrm{L})Z_{m+2}(\mathrm{O}). \tag*{(17)}%
\end{align}

When the $m$th level site is in the L state the partial partition function can
be written as:
\begin{align}
Z_{m}(\mathrm{L})  &  =[Z_{m+1}(\mathrm{R})+wZ_{m+1}(\mathrm{L})]\{Z_{m+1}%
(\mathrm{I})Z_{m+2}(\mathrm{I})+w^{2}w_{\mathrm{p}}w_{\mathrm{h}}%
Z_{m+1}(\mathrm{O})Z_{m+2}(\mathrm{O})\}\nonumber\\
&  +w_{\mathrm{p}}\left[  wZ_{m+1}^{2}(\mathrm{L})Z_{m+2}(\mathrm{R}%
)+Z_{m+1}^{2}(\mathrm{R})Z_{m+2}(\mathrm{L})\right] \nonumber\\
&  +\left[  Z_{m+2}(\mathrm{R})+wZ_{m+2}(\mathrm{L})\right]  wZ_{m+1}%
(\mathrm{I})Z_{m+1}(\mathrm{O}). \tag*{(18)}%
\end{align}

The relation for the R state is obtained from $Z_{m}(\mathrm{L})$ by the
interchange L$\longleftrightarrow$R:
\begin{align}
Z_{m}(\mathrm{R})  &  =[Z_{m+1}(\mathrm{L})+wZ_{m+1}(\mathrm{R})]\{Z_{m+1}%
(\mathrm{I})Z_{m+2}(\mathrm{I})+w^{2}w_{\mathrm{p}}w_{\mathrm{h}}%
Z_{m+1}(\mathrm{O})Z_{m+2}(\mathrm{O})\}\nonumber\\
&  +w_{\mathrm{p}}\left[  wZ_{m+1}^{2}(\mathrm{R})Z_{m+2}(\mathrm{L}%
)+Z_{m+1}^{2}(\mathrm{L})Z_{m+2}(\mathrm{R})\right] \nonumber\\
&  +\left[  Z_{m+2}(\mathrm{L})+wZ_{m+2}(\mathrm{R})\right]  wZ_{m+1}%
(\mathrm{I})Z_{m+1}(\mathrm{O}). \tag*{(19)}%
\end{align}

It is possible to write analogous relations for $Z_{m+1}(\mathrm{\alpha})$ by
properly substituting $m\rightarrow m+1,$ $m+1\rightarrow m+2$ and
$m+2\rightarrow m+3.$ We introduce the following ratios between partial
partition functions at even and odd levels of the lattice:
\begin{align}
x_{m}(\mathrm{I})  &  =Z_{m}(\mathrm{I})/\left(  Z_{m}(\mathrm{L}%
)+Z_{m}(\mathrm{R})\right)  ,\nonumber\\
x_{m}(\mathrm{O})  &  =Z_{m}(\mathrm{O})/\left(  Z_{m}(\mathrm{L}%
)+Z_{m}(\mathrm{R})\right)  ,\nonumber\\
x_{m}(\mathrm{L})  &  =Z_{m}(\mathrm{L})/\left(  Z_{m}(\mathrm{L}%
)+Z_{m}(\mathrm{R})\right)  ,\nonumber\\
x_{m}(\mathrm{R})  &  =1-x_{m}(\mathrm{L}). \tag*{(20)}%
\end{align}

As one moves from a level that is infinitely far away from the origin towards
the origin itself, the recursion relations (16)-(19) will approach fix-point
(FP) solutions, $x_{m}(\alpha)\rightarrow x^{\ast}(\alpha)$, $x_{m+1}%
(\alpha)\rightarrow x^{\ast\ast}(\alpha)$, etc., where $\alpha=$ I, O, L or R.
These fix-point solutions of the recursion relations describe the behavior in
the interior of the Husimi tree. Once the fixed point is reached, the value of
$x^{\ast}(\alpha)$ and $x^{\ast\ast}(\alpha)$ becomes independent of $m.$\ On
a Husimi cactus, a site can be classified as a simultaneous peak and a base
site, or a simultaneous peak and a middle site, depending on the pair of
squares which share the site. Thus, it is expected that the most general FP
solutions will correspond to a 2-cycle solution in which $\ x_{m}(\alpha)$ and
$x_{m+2}(\alpha)\;$tend to the same limit. In this case, we obtain a
sublattice structure in which sites with even levels are different from sites
with odd levels. We can write in this case:
\begin{align}
x_{m}(\mathrm{I})  &  =x_{m+2}(\mathrm{I})=i_{\mathrm{a}},\;x_{m}%
(\mathrm{O})=x_{m+2}(\mathrm{O})=o_{\mathrm{a}},\;x_{m}(\mathrm{L}%
)=x_{m+2}(\mathrm{L})=l_{\mathrm{a}},\nonumber\\
x_{m}(\mathrm{R})  &  =x_{m+2}(\mathrm{R})=1-x_{m}(\mathrm{L})=1-l_{\mathrm{a}%
},\nonumber\\
x_{m+1}(\mathrm{I})  &  =x_{m+3}(\mathrm{I})=i_{\mathrm{b}},\;x_{m+1}%
(\mathrm{O})=x_{m+3}(\mathrm{O})=o_{\mathrm{b}},\;x_{m+1}(\mathrm{L}%
)=x_{m+3}(\mathrm{L})=l_{\mathrm{b}},\nonumber\\
x_{m+1}(\mathrm{R})  &  =x_{m+3}(\mathrm{R})=1-x_{m+1}(\mathrm{L}%
)=1-l_{\mathrm{b}}. \tag*{(21)}\label{eqn21}%
\end{align}

The indices a and b refer to even and odd levels, respectively. Using (21), it
is easy to prove that the system of equations (16) to (19) can be written in
the following form:
\begin{align}
i_{\mathrm{a}}Q_{\mathrm{LR}}  &  =w\left(  1-l_{\mathrm{b}}\right)
l_{\mathrm{b}}i_{\mathrm{a}}+w^{2}w_{\mathrm{p}}w_{\mathrm{h}}o_{\mathrm{b}%
}\left[  l_{\mathrm{b}}(1-l_{\mathrm{a}})+l_{\mathrm{a}}(1-l_{\mathrm{b}%
})\right]  ,\tag*{(22)}\label{eqn22}\\
o_{\mathrm{a}}Q_{\mathrm{LR}}  &  =i_{\mathrm{b}}^{2}i_{\mathrm{a}%
}+i_{\mathrm{b}}\left[  l_{\mathrm{b}}(1-l_{\mathrm{a}})+l_{\mathrm{a}%
}(1-l_{\mathrm{b}})\right]  +w(1-l_{\mathrm{b}})l_{\mathrm{b}}o_{\mathrm{a}%
},\tag*{(23)}\label{eqn23}\\
l_{\mathrm{a}}Q_{\mathrm{LR}}  &  =(1-l_{\mathrm{b}}+wl_{\mathrm{b}%
})[i_{\mathrm{b}}i_{\mathrm{a}}+w^{2}w_{\mathrm{p}}w_{\mathrm{h}}%
o_{\mathrm{b}}o_{\mathrm{a}}]\nonumber\\
&  +wo_{\mathrm{b}}i_{\mathrm{b}}(1-l_{\mathrm{a}}+wl_{\mathrm{a}%
})+w_{\mathrm{p}}(1-l_{\mathrm{b}})^{2}l_{\mathrm{a}}+ww_{\mathrm{p}%
}l_{\mathrm{b}}^{2}(1-l_{\mathrm{a}}),\tag*{(24)}\\
i_{\mathrm{b}}Q_{\mathrm{LR}}^{^{\prime}}  &  =w\left(  1-l_{\mathrm{a}%
}\right)  l_{\mathrm{a}}i_{\mathrm{b}}+w^{2}w_{\mathrm{p}}w_{\mathrm{h}%
}o_{\mathrm{a}}\left[  l_{\mathrm{a}}(1-l_{\mathrm{b}})+l_{\mathrm{b}%
}(1-l_{\mathrm{a}})\right]  ,\tag*{(25)}\label{eqn25}\\
o_{\mathrm{b}}Q_{\mathrm{LR}}^{^{\prime}}  &  =i_{\mathrm{a}}^{2}%
i_{\mathrm{b}}+i_{\mathrm{a}}\left[  l_{\mathrm{a}}(1-l_{\mathrm{b}%
})+l_{\mathrm{b}}(1-l_{\mathrm{a}})\right]  +w(1-l_{\mathrm{a}})l_{\mathrm{a}%
}o_{\mathrm{b}},\tag*{(26)}\label{eqn26}\\
l_{\mathrm{b}}Q_{\mathrm{LR}}  &  =(1-l_{\mathrm{a}}+wl_{\mathrm{a}%
})[i_{\mathrm{a}}i_{\mathrm{b}}+w^{2}w_{\mathrm{p}}w_{\mathrm{h}}%
o_{\mathrm{a}}o_{\mathrm{b}}]\nonumber\\
&  +wo_{\mathrm{a}}i_{\mathrm{a}}(1-l_{\mathrm{b}}+wl_{\mathrm{b}%
})+w_{\mathrm{p}}(1-l_{\mathrm{a}})^{2}l_{\mathrm{b}}+ww_{\mathrm{p}%
}l_{\mathrm{a}}^{2}(1-l_{\mathrm{b}}), \tag*{(27)}\label{eqn27}%
\end{align}
where $Q_{\mathrm{LR}}^{^{\prime}}$ is obtained from $Q_{\mathrm{LR}}$\ by
exchanging a and b subscripts and $Q_{\mathrm{LR}}$\ can be written as:
\begin{equation}
Q_{\mathrm{LR}}=\left(  1+w\right)  \{i_{\mathrm{b}}i_{\mathrm{a}}%
+w^{2}w_{\mathrm{p}}w_{\mathrm{h}}o_{\mathrm{b}}o_{\mathrm{a}}+wi_{\mathrm{b}%
}o_{\mathrm{b}}+w_{\mathrm{p}}[l_{\mathrm{b}}(1-l_{\mathrm{a}})^{2}%
+l_{\mathrm{a}}(1-l_{\mathrm{b}})^{2}]\}. \tag*{(28)}\label{eqn28}%
\end{equation}

\subsection{2-cycle Free Energy}

In order to determine which phase is the stable one at some temperature, we
have to find the free energy of all the possible phases of the system as a
function of $w$. We follow the treatment by Gujrati [36], and provide its
trivial extension to the 2-cycle FP solution shown above. The free energy per
site at the origin of the lattice can be easily calculated from the
expressions for the total partition function $Z$ at the $(m=0)$th, $(m=1)$th
and $(m=2)$th levels.

The total partition function of the system $Z_{0}$ can be written by
considering the two $(m=0)$th branches $\mathcal{C}_{0}\;$meeting at the
origin at the $(m=0)$th level. For this, we need to consider all the possible
configurations in the two branches. This is done by considering all the
configurations that the polymer chain can assume in the two squares that meet
at the origin of the tree. All the possible configurations of the $(m=0) $th
level site (in this case we are not interested in the state of the $(m=1) $th
level sites) are shown in Fig. 7.%
\begin{figure}
[tb]
\begin{center}
\includegraphics[width=3in]
{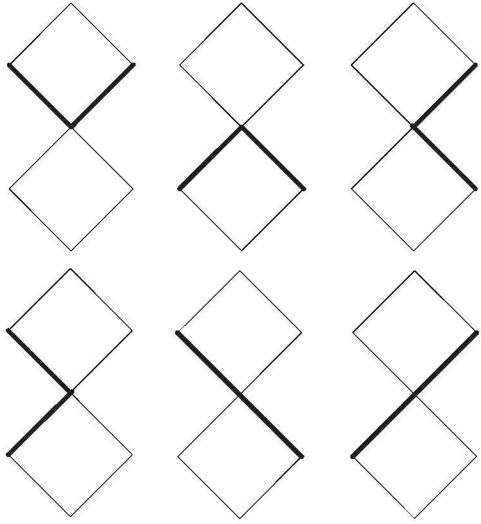}%
\caption{Possible configurations of the polymer chain at the origin of the
tree.}%
\end{center}
\end{figure}
Each of the first two configurations contributes
\[
Z_{0}(\mathrm{I})Z_{0}(\mathrm{O})
\]
to the total partition function. The third and fourth configurations both
contribute
\[
(1/w)Z_{0,\mathrm{g}}(\mathrm{L})Z_{0,\mathrm{g}}(\mathrm{R}),
\]
where the factor $(1/w)$ is needed in order not to take into account the
Boltzmann weight at the origin twice and the subscripts ''g'' and ''t'' refer
to the gauche and trans part of the partition function for L and R states. In
fact, it is possible to separate these two contributions to the partition
function at any level. The ''gauche'' portion is the one corresponding to
configurations such that there is a bending at the $m$th level site, while the
''trans'' portion is the one corresponding to configurations in which the two
bonds coming out of the $m$th level site that we are considering are straight.
It is easily seen that:
\begin{align}
Z_{m,\mathrm{t}}(\mathrm{L})  &  =Z_{m+1}(\mathrm{R})\{Z_{m+1}(\mathrm{I}%
)Z_{m+2}(\mathrm{I})+w^{2}w_{\mathrm{p}}w_{\mathrm{h}}Z_{m+1}(\mathrm{O}%
)Z_{m+2}(\mathrm{O})\nonumber\\
&  +w_{\mathrm{p}}Z_{m+1}(\mathrm{R})Z_{m+2}(\mathrm{L})\}+wZ_{m+2}%
(\mathrm{R})Z_{m+1}(\mathrm{I})Z_{m+1}(\mathrm{O}), \tag*{(29)}%
\end{align}
and
\begin{align}
Z_{m,\mathrm{g}}(\mathrm{L})  &  =wZ_{m+1}(\mathrm{L})\{Z_{m+1}(\mathrm{I}%
)Z_{m+2}(\mathrm{I})+w^{2}w_{\mathrm{p}}w_{\mathrm{h}}Z_{m+1}(\mathrm{O}%
)Z_{m+2}(\mathrm{O})\nonumber\\
&  +w_{\mathrm{p}}Z_{m+1}(\mathrm{L})Z_{m+2}(\mathrm{R})\}+w^{2}%
Z_{m+2}(\mathrm{L})Z_{m+1}(\mathrm{I})Z_{m+1}(\mathrm{O}). \tag*{(30)}%
\end{align}

Finally, the fifth and sixth configurations contribute $Z_{0,\mathrm{t}}%
^{2}(\mathrm{L})$ and $Z_{0,\mathrm{t}}^{2}(\mathrm{R})$, respectively. It is
then possible to write:
\begin{equation}
Z_{0}=2Z_{0}(\mathrm{I})Z_{0}(\mathrm{O})+(2/w)Z_{0,\mathrm{g}}(\mathrm{L}%
)Z_{0,\mathrm{g}}(\mathrm{R})+Z_{0,\mathrm{t}}^{2}(\mathrm{L})+Z_{0,\mathrm{t}%
}^{2}(\mathrm{R}). \tag*{(31)}\label{eqn31}%
\end{equation}

It is clear that $Z_{0}$ is the total partition function of the system
obtained by joining two branches $\mathcal{C}_{0}$ together at the origin.
Now, let us imagine taking away from the lattice the two squares that meet at
the origin. This leaves behind four different branches $\mathcal{C}_{1}$ and
two branches $\mathcal{C}_{2}$. We connect the two $\mathcal{C}_{2}$ branches
to form a smaller cactus whose partition function is denoted by $Z_{2}$.
Similarly, we join two of the $\mathcal{C}_{1}$ branches to form an
intermediate cactus whose partition function is denoted by $Z_{1}$. We can
form two such intermediate cacti out of the four $\mathcal{C}_{1}$ branches.
Each partition function $Z_{1}$ or $Z_{2}$ can be written in a form that is
identical to that of Equation (31):
\begin{equation}
Z_{i}=2Z_{i}(\mathrm{I})Z_{i}(\mathrm{O})+(2/w)Z_{i,\mathrm{g}}(\mathrm{L}%
)Z_{i,\mathrm{g}}(\mathrm{R})+Z_{i,\mathrm{t}}^{2}(\mathrm{L})+Z_{i,\mathrm{t}%
}^{2}(\mathrm{R}), \tag*{(32)}\label{eqn32}%
\end{equation}
where $i=1,2.$

The difference between the free energy of the complete cactus and that of the
three reduced cacti is just the free energy corresponding to a pair of squares
so that, following Gujrati [36], we can write the adimensional free energy
\textit{per square} (without the conventional minus sign) as:
\begin{equation}
\omega\equiv\omega_{\text{sq}}=%
\frac12
\ln[Z_{0}/\{Z_{1}^{2}Z_{2}\}]. \tag*{(33)}\label{eqn33}%
\end{equation}

It is possible to write:
\begin{equation}
Z_{0}=B_{0}^{2}Q_{2}(i_{\mathrm{a}},o_{\mathrm{a}},l_{\mathrm{a}%
},i_{\mathrm{b}},o_{\mathrm{b}},l_{\mathrm{b}}), \tag*{(34)}\label{eqn34}%
\end{equation}%
\begin{equation}
Z_{1}=B_{1}^{2}Q_{2}^{^{\prime}}(i_{\mathrm{a}},o_{\mathrm{a}},l_{\mathrm{a}%
},i_{\mathrm{b}},o_{\mathrm{b}},l_{\mathrm{b}}), \tag*{(35)}\label{eqn35}%
\end{equation}%
\begin{equation}
Z_{2}=B_{2}^{2}Q_{2}(i_{\mathrm{a}},o_{\mathrm{a}},l_{\mathrm{a}%
},i_{\mathrm{b}},o_{\mathrm{b}},l_{\mathrm{b}}), \tag*{(36)}\label{eqn36}%
\end{equation}
where we have introduced
\begin{equation}
B_{m}=Z_{m}(\mathrm{L})+Z_{m}(\mathrm{R}), \tag*{(37)}\label{eqn37}%
\end{equation}
and where $Q_{2}$ is the following polynomial of $i_{\mathrm{a}}%
,o_{\mathrm{a}},l_{\mathrm{a}},i_{\mathrm{b}},o_{\mathrm{b}}$ and
$l_{\mathrm{b}}$:
\begin{equation}
Q_{2}=2i_{\mathrm{a}}o_{\mathrm{a}}+(2/w)l_{\mathrm{a,g}}(1-l_{\mathrm{a}%
})_{\text{\textrm{g}}}+l_{\mathrm{a,t}}^{2}+(1-l_{\mathrm{a}})_{\mathrm{t}%
}^{2}; \tag*{(38)}\label{eqn38}%
\end{equation}
$l_{\mathrm{a,t}}$ and $l_{\mathrm{a,g}}$\ correspond to the trans and gauche
portions of $l_{\mathrm{a}}$, respectively, and $Q_{2}^{^{\prime}}$\ \bigskip
is obtained from $Q_{2}$ by interchanging a and b subscripts.

It is also easily seen that:
\begin{equation}
B_{0}=B_{1}^{2}B_{2}Q_{\mathrm{LR}}(i_{\mathrm{a}},o_{\mathrm{a}%
},l_{\mathrm{a}},i_{\mathrm{b}},o_{\mathrm{b}},l_{\mathrm{b}}). \tag*{(39)}%
\label{eqn39}%
\end{equation}
so that the free energy per square can be written as:%

\begin{equation}
\omega\equiv\omega_{\text{sq}}=\ln\left(  \frac{Q_{\mathrm{LR}}}%
{Q_{2}^{^{\prime}}}\right)  . \tag*{(40a)}%
\end{equation}
The free energy per site $\omega_{\text{site}}$ is proportional to
$\omega_{\text{sq}}:$%

\begin{equation}
\omega_{\text{site}}\equiv\omega_{\text{sq}}/2, \tag*{(40b)}%
\end{equation}
since there are two sites per square.

The usual Helmholtz free energy per square can be obtained from $\omega$
through:
\begin{equation}
F=-T\omega. \tag*{(41)}\label{eqn41}%
\end{equation}

\bigskip If it happens that the even and odd sites are not different, we
obtain a 1-cycle FP-solution. Below, we will consider the two solutions separately.

\subsection{1-cycle solution}

In the 1-cycle scheme, we have $x_{m}(\alpha)=x_{m+1}(\alpha)$\ as they
converge to the same fix point. Thus, we have: $i_{\mathrm{a}}=i_{\mathrm{b}%
}=i$, $o_{\mathrm{a}}=o_{\mathrm{b}}=o$ and $l_{\mathrm{a}}=l_{\mathrm{b}%
}=l.\;$In this case the system of equations reduces to:
\begin{align}
iQ_{\mathrm{LR}}  &  =w\left(  1-l\right)  li+2w^{2}w_{\mathrm{p}%
}w_{\mathrm{h}}ol(1-l),\tag*{(42)}\label{eqn42}\\
oQ_{\mathrm{LR}}  &  =i^{3}+2il(1-l)+w(1-l)lo,\tag*{(43)}\label{eqn43}\\
lQ_{\mathrm{LR}}  &  =\left[  (1-l)+wl\right]  \left\{  i^{2}+w^{2}%
w_{\mathrm{p}}w_{\mathrm{h}}o^{2}+w_{\mathrm{p}}l(1-l)+wio\right\}  ,
\tag*{(44)}\label{eqn44}%
\end{align}
with:
\begin{equation}
Q_{\mathrm{LR}}=(1+w)\left\{  i^{2}+w^{2}w_{\mathrm{p}}w_{\mathrm{h}}%
o^{2}+w_{\mathrm{p}}l(1-l)+wio\right\}  . \tag*{(45)}\label{eqn45}%
\end{equation}
From (44), it is easy to show that we must have $l=%
\frac12
$ for every solution obtained in this scheme.

One solution that exists for every value of $w$ (and, hence, of $T$) is the
following:
\begin{equation}
l=%
\frac12
,o=0,i=0. \tag*{(46)}\label{eqn46}%
\end{equation}
This represents a liquid-like phase without any I, or O states. We label this
phase \textit{metastable liquid} (ML) because, as we will show below, it
\textit{never} represents the equilibrium phase of the system even though it
exists at all temperatures. For large enough $w$, there is another solution of
the system of equations with nonzero values for $o$ and $i$, and it has to be
found numerically. We label this second liquid-like phase \textit{equilibrium
liquid} (EL) since, as our free energy calculations will show, it represents
the equilibrium phase of the system at high temperature. The temperature at
which EL appears is a function of $w_{\mathrm{p}}$ and $w_{\mathrm{h}}$. We
will call the temperature $T_{\mathrm{MC}}$, at which the equilibrium liquid
appears, the mode coupling temperature for reasons that will become clear below.

Table 1 shows how the value of $T_{\mathrm{MC}}$ changes as a function of $a$
(both positive and negative values of $a$ are considered, see below) as we
keep $b$ equal to zero.

\begin{center}%
\begin{tabular}
[c]{|c|c|}\hline
a & $T_{\text{MC}}$\\\hline
-1 & 2.9586\\\hline
-0.8 & 2.6183\\\hline
-0.5 & 2.1876\\\hline
-0.2 & 1.7359\\\hline
0 & 1.4427\\\hline
0.2 & 1.1600\\\hline
0.5 & 0.7653\\\hline
0.8 & 0.4156\\\hline
1 & 0\\\hline
\end{tabular}

Table 1: Values of $T_{\mathrm{MC}}$ as a function of $a$ (with $b=0$).
\end{center}

For the ML, we have $o=i=0,l=%
\frac12
$ so that:
\begin{equation}
Q_{2}=\frac{1}{2(1+w)}, \tag*{(47)}\label{eqn47}%
\end{equation}
and
\begin{equation}
Q_{\mathrm{LR}}=\frac{w_{\mathrm{p}}(1+w)}{4}, \tag*{(48)}\label{eqn48}%
\end{equation}
so that the ML free energy per square assumes the simple form:
\begin{equation}
\omega_{\text{ML}}=\ln\left[  w_{\mathrm{p}}\left(  1+w\right)  ^{2}/2\right]
=\ln\left[  \left(  1+w\right)  ^{2}/2\right]  +\ln(w_{\mathrm{p}}),
\tag*{(49a)}\label{eqn49}%
\end{equation}
while, for the EL, we have to substitute the numerical solutions $i(w),$
$o(w)$, and $l=%
\frac12
$ obtained from Eqs. (42)-(45) in the expression for the free energy. The
ML\ entropy per square is given by%

\begin{equation}
S_{\text{ML}}=\ln\left[  \left(  1+w\right)  ^{2}/2\right]  +2w/[T(1+w).
\tag*{(49b)}%
\end{equation}
The corresponding energy is given by%

\begin{equation}
E_{\text{ML}}=2w/(1+w)+a. \tag*{(49c)}%
\end{equation}
It is easily seen that the ML specific heat is always non-negative. It is very
important to observe that the free energy of the ML phase does not depend on
the value of the parameter $a$ (except for the additive factor $\ln
(w_{\mathrm{p}})$) while the free energy of EL strongly does. At absolute
zero, the ML entropy and energy go to ln$(1/2)$ , and $a$, respectively. We
find the \textit{modified free energy }%

\begin{equation}
\overset{\sim}{F}=F-\varepsilon_{\mathrm{p}} \tag*{(50)}%
\end{equation}
more convenient to use than $F$ itself since, at $T=0$, the ground state is
the one in which all the bonds are parallel to each other and the free energy
of the system is equal to $\varepsilon_{\mathrm{p}}$ so that the crystalline
ground state has always $\overset{\sim}{F}=0,$\ regardless of the value of
$\varepsilon_{\mathrm{p}}$. The free energy curves for the EL and ML phases
are shown in Fig. 8. We immediately observe that ML at very low temperatures
$T\lesssim0.48$ (dash-double dot line in Fig. 8 originating at the
origin)\ has \textit{negative entropy}, since its free energy $\overset{\sim
}{F}$ is increasing with the temperature. A negative entropy is not possible
for states that can exist in Nature.%
\begin{figure}
[tb]
\begin{center}
\includegraphics[width=4.5in]
{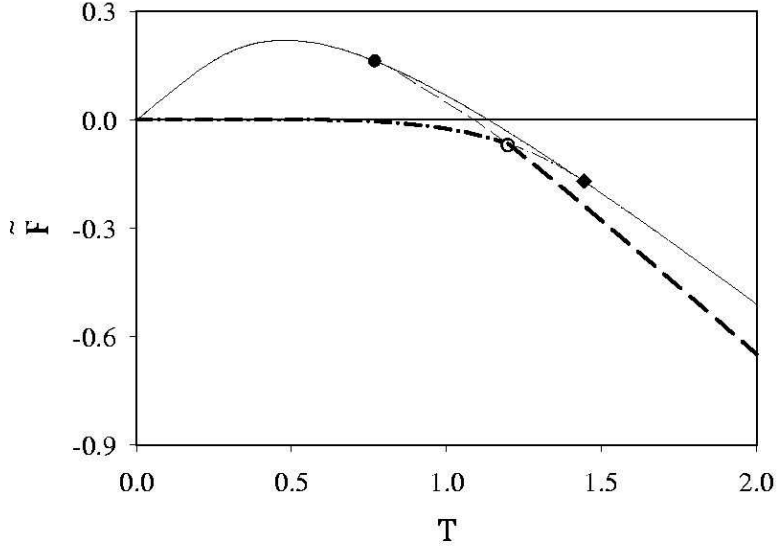}%
\caption{$a=0.5,b=0$. Free energy in the 2-cycle FP scheme for the ML
(continuous line), EL/SCL (dashed line) and CR (dash-dot line). We also show
$T_{\mathrm{CRE}}(\blacklozenge)$, $T_{\mathrm{M}}(\circ)$ and $T_{\mathrm{MC}%
}(\bullet).$ Here, as well as in Figures 10 to 12, the stable phases are
represented by thick lines while the metastable phases are represented by thin
lines.}%
\end{center}
\end{figure}

\subsection{2-cycle solution}

The phase diagram obtained in the 1-cycle solution scheme \textit{cannot} be
complete because, at low temperatures, ML cannot be the stable phase. At
$T=0,$ CR contains an alternating ordered sequence of L and R states in
addition to having $l=%
\frac12
$ and no O and I states; see Fig. 4. This is a 2-cycle pattern in L and R that
is completely missed by the 1-cycle calculation performed above. For ML, $l=%
\frac12
$ also, but L and R are statistically distributed. One of these distributions
must be the crystal state at $T=0$; indeed, $F_{\mathrm{ML}}%
(T=0)=F_{\mathrm{CR}}(T=0).$ Despite this, ML immediately above $T=0$ can not
represent CR, as it has negative entropy. To obtain the alternating sequence
in CR at $T>0$, the above 1-cycle FP scheme is not sufficient to completely
describe the physics of the system. We also observe that for $T>0$, there must
be local Gujrati-Goldstein excitations [10,11] creating imperfections by local
L$\longleftrightarrow$R interchanges in the ordered [..LRLRLR..] sequence. The
excitations change a local string LRL into LLL, or RLR into RRR within a
square and require 4 bends only. Other excitations, which require (L or
R)$\longleftrightarrow$(I or O) on the cactus, can not be done locally and
require infinite amount of energy, and need not be considered. This means that
the local density $l$ or $r$ will no longer be
${\frac12}$%
. However, if $l>%
\frac12
$ at some site, then $r>%
\frac12
$ at the next site, followed by $l>%
\frac12
$ on the next site and so on.

There are three solutions for the complete system of Eqs. (22)-(27) for any
given value of the weights $w$, $w_{\mathrm{p}}$ and $w_{\mathrm{h}}$:

i. A metastable liquid ML (already found in the 1-cycle FP scheme) with:

$l_{\mathrm{a}}=l_{\mathrm{b}}=r_{\mathrm{a}}=r_{\mathrm{b}}=%
\frac12
$ and $i_{\mathrm{a}}=i_{\mathrm{b}}=o_{\mathrm{a}}=o_{\mathrm{b}}=0$.

As seen above, this phase represents a liquid phase in which no O and I states
are present. The R and L states are randomly distributed in the lattice with
the only constraint of having the same number of L and R states at both odd
and even layers. This solution exists for any temperature and its free energy
has a maximum at $T=T_{\mathrm{K}}\simeq0.48$.

ii. An equilibrium liquid EL characterized by the presence of all possible
states I, O, L and R at both odd and even levels, so that

$l_{\mathrm{a}}=l_{\mathrm{b}}=r_{\mathrm{a}}=r_{\mathrm{b}}=%
\frac12
$ and $i_{\mathrm{a}}$, $i_{\mathrm{b}}$, $o_{\mathrm{a}}$, $o_{\mathrm{b}%
}\neq0$.

In the 2-cycle solution, $i$ and $o$ on the two sublattices are different,
which makes this solution different from the 1-cycle EL\ solution, in which
there is no sublattice structure. Despite this, EL phases in both schemes
have\textit{\ identical} free energy and various densities. Thus, we no longer
make any distinction between the two solutions and identify both of them as
the same EL phase. As seen in the previous subsection, the free energy of this
phase depends on the value of the parameters $a$ and $b$. This solution exists
only for temperatures larger than $T=T_{\mathrm{MC}}(a,b)$.

iii. A crystal phase CR with double degeneracy that is the ground state and
that exists for temperatures lower than $T=T_{\mathrm{CRE}}=1/\ln(2)$. The
state is perfectly ordered at zero temperature and disorders as the
temperature is raised. Fig. 9 shows how the values of $l_{a}$ and $l_{b}$
change with temperature for the CR phase: the two degenerate solutions
correspond to a different labeling of the lattice sites where the odd and even
levels are just exchanged with each other.%
\begin{figure}
[tb]
\begin{center}
\includegraphics[width=4.5in]
{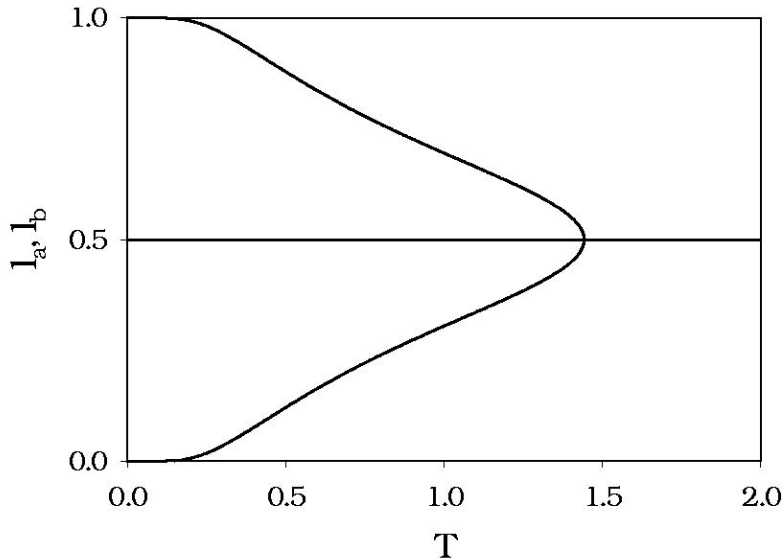}%
\caption{Dependence of $l_{\mathrm{a}}$ and $l_{\mathrm{b}}$ on the
temperature for the two phases obtained at low temperature in the 2-cycle
fixed point scheme.}%
\end{center}
\end{figure}

The solutions of the system of Eqs. (22)-(27) corresponding to CR and to ML do
\textit{not} depend on the strength of the three- and four- site interaction
and, therefore, the free energy curves corresponding to these two phases do
not change when the parameters $a$ and $b$ change. In contrast, the free
energy of the EL phase depends on the value of $a$ and $b$.

For $a<1$, we have the ground state at $T=0$ in which all the bonds are
parallel and there are no bends in the systems. If $a>1$, the four-site
interaction becomes more important than the three-site interaction and the
ground state at $T=0$ is a step like walk in which there are no parallel bonds
and all the sites of the lattice are in a gauche conformation (so that
$N_{\mathrm{g}}=N_{\text{S}}$, as said earlier). The two possible ground
states are shown in Fig. 4.

\section{RESULTS AND DISCUSSION}

\subsection{Thermodynamic Functions}

\subsubsection{$b=0$}

The complete free energy diagram for $a=0.5$ is given in Fig. 8. The
equilibrium phases are represented by the disordered EL at high temperatures
and the ordered CR at low temperatures, with a first order transition at a
temperature $T_{\mathrm{M}}$ between the two phases, with a discontinuity in
the first derivative of the free energy with temperature. This remains true as
long as $a>0$. The situation with $a\leq0$ is different and is discussed later.

The existence of a discontinuous melting temperature for $a>0$ makes it
possible to have a supercooled liquid phase through continuation. For
$T>T_{\mathrm{M}}$, the EL phase is the stable one. If the liquid phase is
cooled in such a way that it is \textit{not }allowed to undergo the melting
transition at $T_{\mathrm{M}}$, then it is possible to have (for
$T<T_{\mathrm{M}}$) a supercooled liquid (SCL). The free energy of SCL\ is
obtained by\textit{\ continuing} the free energy of the EL\ phase. This free
energy meets critically (i. e. with continuous slopes) with the ML free energy
at a temperature that, as before, we call $T_{\mathrm{MC}}$. For $a=0.5$ we
find that $T_{\mathrm{MC}}>T_{\mathrm{K}}$, where $T_{\mathrm{K}}$ is the
temperature where the ML free energy has its maximum. The critical transition
between ML and SCL is a \textit{liquid-liquid transition} between two liquid
phases. As $a$ increases, $T_{\mathrm{MC}}$ moves towards $T_{\mathrm{K}}$. We
have observed that for $a\gtrsim0.8,$ $T_{\mathrm{MC}}<T_{\mathrm{K}}$
(results not shown). In particular, the EL free energy curve itself has a
maximum in this case before it merges with the ML curve and, consequently, has
an \textit{unphysical} portion corresponding to the entropy crisis below its maximum.

Figs. 10 and 11 show the entropy and the specific heat vs. temperature,
respectively, corresponding to the free energy results shown in Fig. 8. As
explained before in the case of the free energy-temperature graphs, the curves
corresponding to CR and to ML do not depend on the choice of $a$ and $b$.
Table 2 shows how the value of $T_{\mathrm{M}}$ changes as a function of $a$
(in the case $b=0$).

\begin{center}%
\begin{tabular}
[c]{|c|c|}\hline
$a$ & $T_{\text{M}}$\\\hline
0 & 1.443\\\hline
0.2 & 1.351\\\hline
0.5 & 1.198\\\hline
0.8 & 1.009\\\hline
1 & 0.878\\\hline
\end{tabular}

Table 2: Values of $T_{\mathrm{M}}$ as a function of $a$ (with $b=0$).
\end{center}

Only positive values of $a$ are considered since, as it will be shown below,
when the parameter $a$ is negative, there is no first-order melting in the
system, provided that $b=0$.%
\begin{figure}
[t]
\begin{center}
\includegraphics[height=4in]
{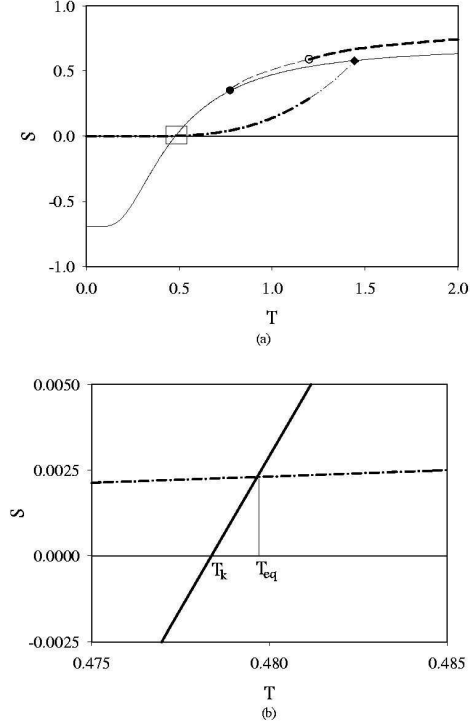}%
\caption{$a=0.5,b=0$. (a) Entropy in the 2-cycle FP scheme for the ML
(continuous line), EL/SCL (dashed line) and CR (dash-dot line). We also show
$T_{\mathrm{CRE}}(\blacklozenge)$, $T_{\mathrm{M}}(\circ)$ and $T_{\mathrm{MC}%
}(\bullet)$; (b) magnification of the area contained in the box in (a).}%
\end{center}
\end{figure}

We can calculate the density of gauche bonds $g$ and the density of pairs of
parallel bonds $p$ as a function of $T$ and $a$. We can write:
\begin{equation}
g=\partial\omega_{\text{site}}/\partial(\ln(w))|_{w_{\mathrm{p}}%
,w_{\mathrm{h}}}, \tag*{(51)}\label{eqn51}%
\end{equation}%
\begin{equation}
p=\partial\omega_{\text{sq}}/\partial(\ln(w_{\mathrm{p}}))|_{w,w_{\mathrm{h}}%
}, \tag*{(52)}\label{eq52}%
\end{equation}
and calculate the two densities from these derivatives. Note that we have
defined the gauche bond density $g$ per site, while the density of parallel
bond pairs $p$ is defined per square, since each square contributes one such
pair in the ideal CR at absolute zero.%
\begin{figure}
[tb]
\begin{center}
\includegraphics[width=4.2in]
{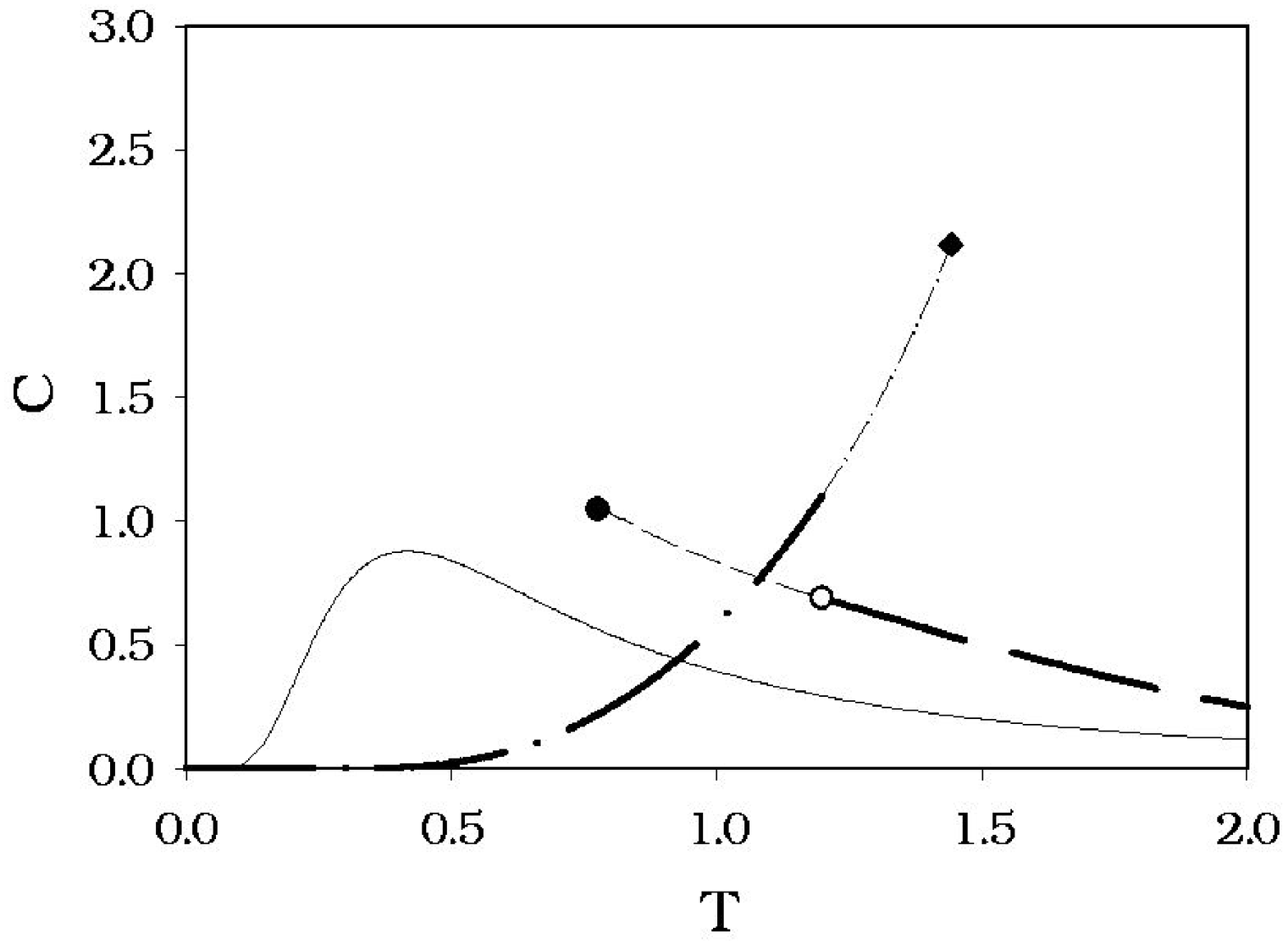}%
\caption{$a=0.5,b=0$. Specific heat in the 2-cycle FP scheme for the ML
(continuous line), EL/SCL (dashed line) and CR (dash-dot line). We also show
$T_{\mathrm{CRE}}(\blacklozenge)$, $T_{\mathrm{M}}(\circ)$ and $T_{\mathrm{MC}%
}(\bullet)$.}%
\end{center}
\end{figure}
Fig. 12 shows the gauche bond density and the parallel bond density in the
case of $a=0.5$, $b=0$.%
\begin{figure}
[tbtb]
\begin{center}
\includegraphics[height=4in]
{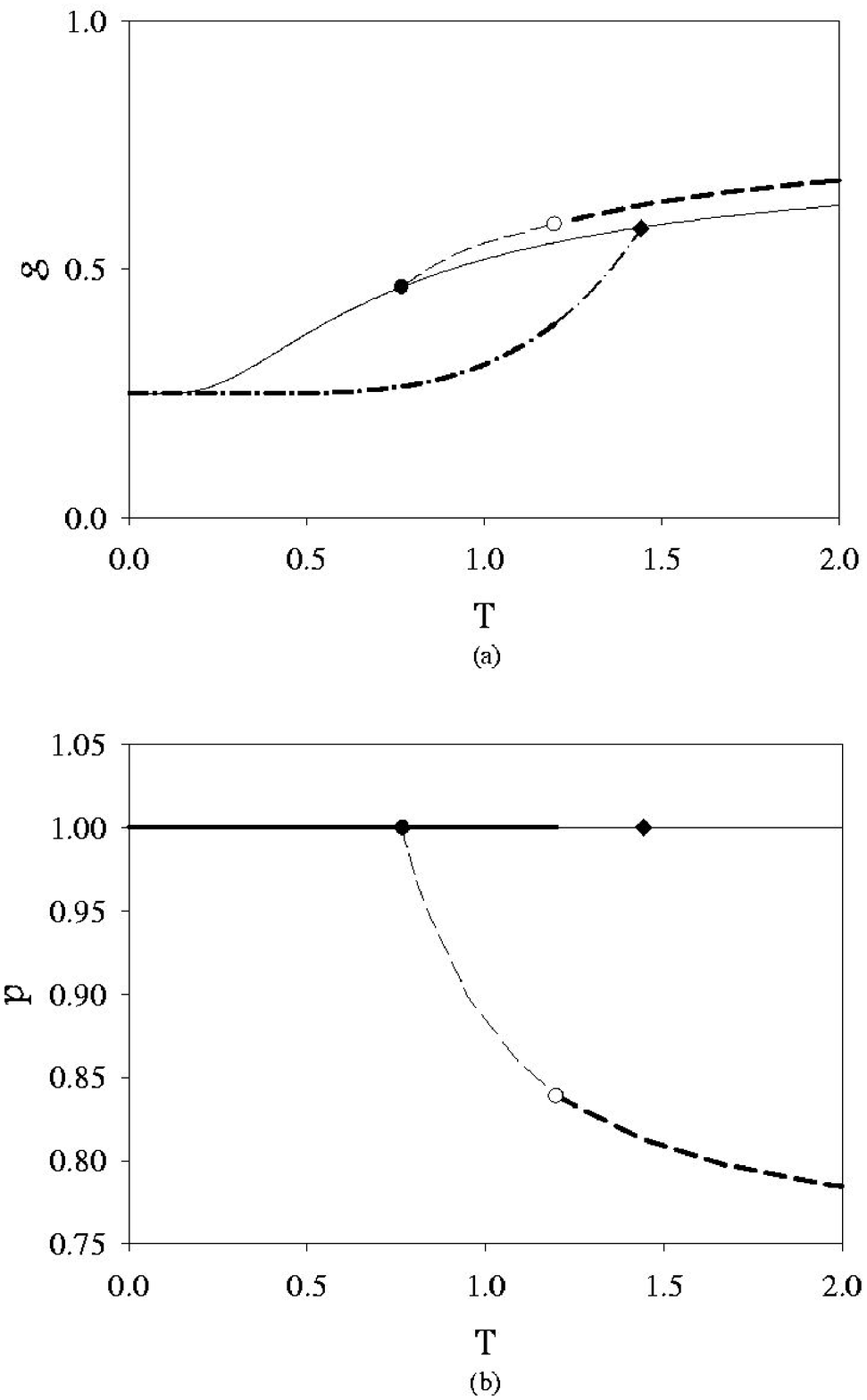}%
\caption{(a) Gauche bond and (b) parallel bond density in the 2-cycle FP
scheme for the ML (continuous line), EL/SCL (dashed line) and CR (dash-dot
line). We also show $T_{\mathrm{CRE}}(\blacklozenge)$, $T_{\mathrm{M}}(\circ)$
and $T_{\mathrm{MC}}(\bullet)$.}%
\end{center}
\end{figure}

\subsubsection{$b\neq0$}

The effect of changing the value of the parameter $b$ is shown in Fig. 13 for
$a=0.5$. As we can see, a change in $b$ does not have any effect on the free
energy of the CR and ML phases but it does affect the EL phase. Apparently,
the effect of $b$ is smaller than the effect of $a$, since the value of the
melting temperature does not change significantly as we change $b$. It is
worth noting that the presence of the hairpin term alone, even in the absence
of the interaction between pairs of parallel bonds, is sufficient to transform
the melting transition from second-order, as seen in the case $a=b=0$
(original Flory model), to first-order.\ If $b$ is negative, the melting and
mode-coupling temperatures decrease. If $b$ is positive, instead, these
temperatures increase and the melting transition turns into a second-order
transition. This is true for any positive value of $b$ in the case $a=0$,
while it is true for large enough negative values of $b$ when $a$ is positive.%
\begin{figure}
[tb]
\begin{center}
\includegraphics[width=4.5in]
{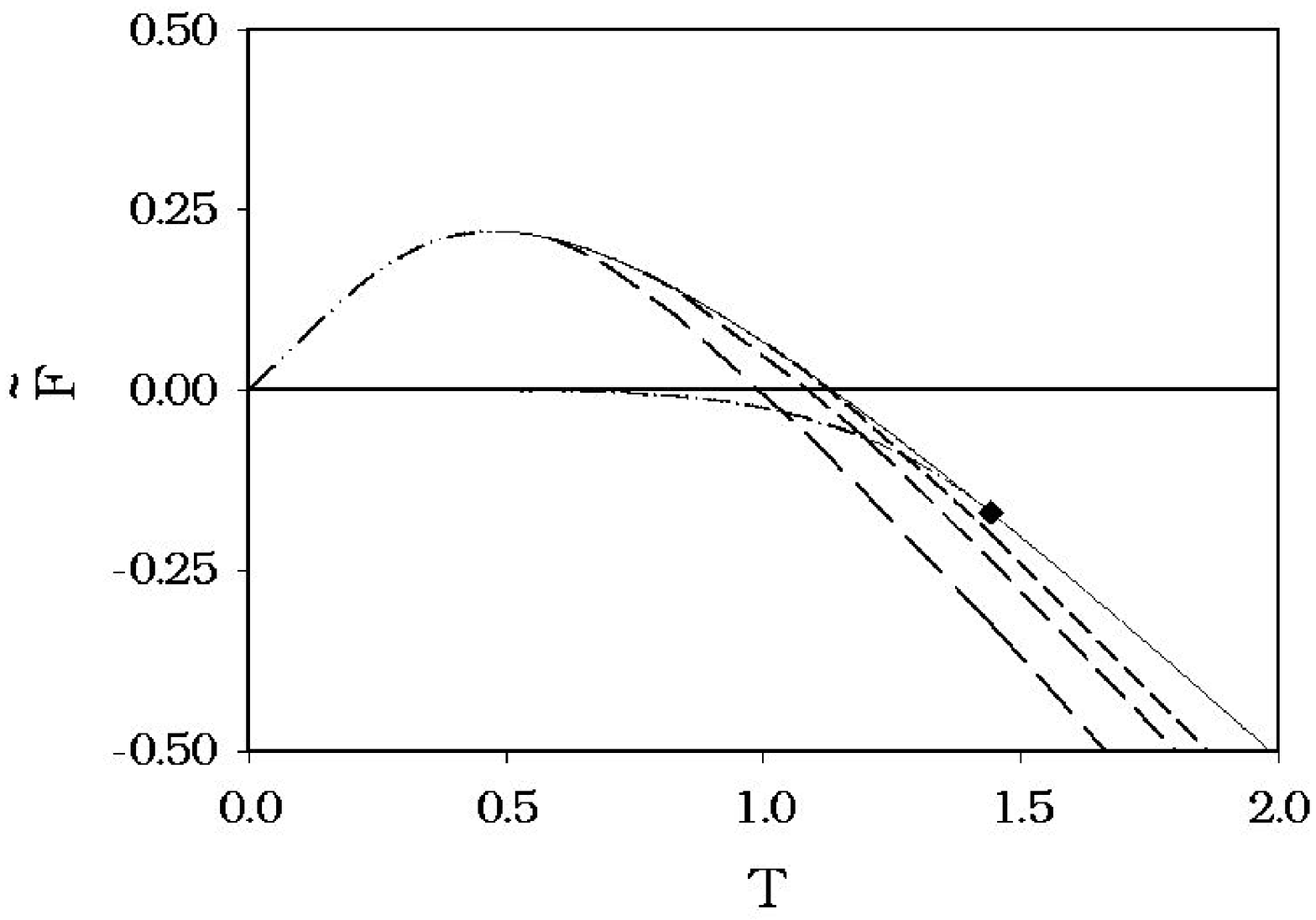}%
\caption{Effect of $b$ on the phase diagram of the system ($a=0.5$). The free
energy of ML and CR\ does not depend on $b$. Three EL/SCL curves are shown
corresponding to $b=-0.4$ (long dash), $b=0$ (medium dash) and $b=0.4$ (short
dash).}%
\end{center}
\end{figure}

It is interesting to observe that the parallel bond density in ML and in CR is
always unity while its value for the EL depends on the temperature.

\subsubsection{$a>1$}

We consider now the case $a>1$. This case corresponds to a ground state that
is \textit{not} crystalline, as shown in Fig. 4(b). When the four-site
interaction is stronger than the three-site interaction, the polymer assumes a
configuration that is such that the number of parallel bonds is minimized. In
this case there is a very high number of gauche conformations at low
temperatures and, even though the polymer assumes an ordered configuration on
the lattice, it is not a crystalline configuration according to our
definition. Therefore, we do not consider this case any further.

\subsubsection{$a<0$}

It is also possible to analyze the case in which $a<0$. When $a$ is negative
(corresponding to a negative four-site interaction energy $\varepsilon
_{\text{p}}$) the temperature $T_{\text{MC}}$ at which the EL appears moves to
higher values. Since the temperature $T_{\text{M}}(\equiv T_{\text{CRE}})$ at
which CR appears is unaffected by the choice of the value of $a$, a shift of
the origin of the EL phase to higher temperatures makes ML the
\textit{equilibrium phase} for temperatures between $T_{\text{M}}$\ and
$T_{\text{MC}}>T_{\text{M}}$; it is no longer a metastable phase in this
range$.$ We identify the equilibrium portion of the ML phase as a new
equilibrium phase, and denote it by EL$_{\text{ML}}$; the subscript is a
reminder that the phase is associated with the ML phase. The new phase is
again a liquid phase. Hence, the transition at $T_{\text{MC}}$ is a liquid
-liquid transition and is continuous. Similarly, the transition at
$T_{\text{M}}$ between CR and EL$_{\text{ML}}$ is also a continuous
transition. Since ML exists below $T_{\text{M}}$, we can formally treat the ML
phase below $T_{\text{M}}$ as obtained by continuation of EL$_{\text{ML}}$
below $T_{\text{M}}$, but this continuation should not be taken as a
supercooled liquid below $T_{\text{M}}$, as there will be no energy barrier
due to the continuous melting transition. Furthermore, since the liquid-liquid
transition at $T_{\text{MC}}$ occurs at a temperature higher than the melting
temperature, this case has no relevance for studying the glass transition.
Hence, we do not pursue it further.\ The same behavior is observed, as
explained above, when $a=0$ and $b>0$.

\subsubsection{Choice of $a$}

Because of the above considerations, we consider $a$ in the range 0
$<$%
$a$
$<$%
1. In this range, the four-site interaction is repulsive (the system spends
some energy to align two segments parallel to each other) and the ground state
is the crystalline one [see Fig. 4(a)]. In this case, the model exhibits a
first order melting transition at a temperature $T_{\text{M}}=T_{\text{M}}(a)$
between EL, that is stable at temperatures higher than $T_{\text{M}}$, and CR,
that is stable for temperatures lower than $T_{\text{M}}$. It can be observed
that the discontinuity in the specific heat at $T_{\text{MC}}$ is a function
of the parameters $a$ and $b$. In particular, if we fix $b$, as $a$ increases
the discontinuity gets smaller and smaller as long as $a$
$<$%
0.8, and then starts growing again while the transition temperature keeps
moving to lower values; the results are not shown.

The crystalline phase is an ordered one but, unlike the ground state predicted
by the original Flory model [1,2] (Fig. 1), it has non zero entropy. It also
satisfies the Gujrati-Goldstein bounds. The I and O states disappear in the
crystalline phase but this phase has non zero entropy because of the many
possible configurations that the polymer chain can assume corresponding to
different sequences of L and R states. The entropy of the crystalline phase
goes to zero only when the temperature goes to zero, which is consistent with
the third law of thermodynamics.

If the cooling process is such that the system can avoid crystallization at
$T_{\text{M}}$, the equilibrium liquid EL can be supercooled to give rise to
SCL that transforms into the metastable liquid ML through a liquid-liquid
second-order transition at $T_{\text{MC}}\leq T_{\text{M}}$ (no latent heat is
associated with the transition). The metastable liquid and the equilibrium
liquid phases are somehow similar. The metastable liquid consists of a random
sequence of R and L states, while the equilibrium liquid consists of a random
sequence of R, L, I and O states. The presence of O and I states makes the
total energy and the entropy of the equilibrium liquid larger than those of
the metastable liquid, see Figs. 10 and 11.

\subsection{Relation with the Mode-coupling Transition}

We tentatively identify the critical temperature $T_{\text{MC}}$ of the
liquid-liquid transition as the mode-coupling transition temperature\ because
the transition exhibits some of the features predicted by the original mode
coupling theory at the critical mode-coupling transition temperature. It
should be remarked that our equilibrium investigation \textit{cannot} provide
any direct information about the dynamics at $T_{\text{MC}}$ except by
inference. Hence, the connection we allude to above should only be taken as
tentative, in view of the fact that the mode-coupling transition, applied
successfully to simple fluids, is considered to be a dynamic transition. We
can only add that the mode-coupling theory is not well-understood for long
polymers, and it is not clear what its predictions might be for infinite
polymers that we are investigating here.

According to this theory, the dynamics slows down according to Eq. (3) near
$T_{\text{MC}}.$ The local molecular structure freezes and only long-time
cooperative jumps are allowed below this temperature. Thus, the dynamic
transition is between two disordered states, very much like the thermodynamic
liquid-liquid transition we observe in our calculation. Let us consider the
behavior of the correlation length $\xi_{\text{SCL}}$ of the system near the
critical temperature $T_{\text{MC}}$. As $T_{\text{MC}}$\ is approached from
above $(T\rightarrow T_{\text{MC}}^{+}),$ the correlation length
$\xi_{\text{SCL}}$ of the supercooled liquid must diverge to infinity because
the transition is continuous. It is very easy to observe from the results that
there is a \textit{discontinuity} in the specific heat of the system at the
transition from the supercooled liquid to the metastable liquid. Indeed, the
SCL terminates at $T_{\text{MC}}$ as $T\rightarrow T_{\text{MC}}^{+}.$ The
disappearance of SCL is what gives rise to this divergence, which will
contribute to the critical slowing down of the system. Such a critical slowing
down is exhibited at the mode-coupling transition [39]; see Eq. (3).

On the other hand, ML exists at all temperatures. Thus, there would be
\textit{no} divergence at $T_{\text{MC}}$ in the correlation length
$\xi_{\text{ML}} $ associated with ML . Indeed, its specific heat remains
continuous. This will imply that the dynamics of the system should not undergo
any significant change at the critical temperature $T_{\text{MC}}$ if we
approach it while heating up ML in such a way that ML is not allowed to turn
into SCL. In simulation, one can investigate ML by suppressing fast
relaxations that are supposed to freeze at the mode-coupling temperature. Such
an attempt has already been made [25] where one sees no anomalous behavior at
the mode-coupling temperature. Parisi and coworkers [25] while analyzing a
Lennard-Jones system have observed this kind of dynamics. In their approach,
the fast dynamics of the system (the one pertinent to the supercooled liquid
in our model) is suppressed and only the slow dynamics of the system is taken
into account. The slow dynamics is described as a relaxation process taking
place in a connected network of potential energy minima. Indeed, the authors
only observes an Arrhenius behavior in the relaxation time, as opposed to the
Vogel-Tammann-Fulcher behavior associated with the mode-coupling transition.
Even though the system studied by Parisi and coworkers is very different from
the polymer system studied here, it is important to note that all the
numerical results obtained in the case of the Lennard-Jones fluid are in
agreement with the experimental findings in non-network forming glasses and
especially in glasses that are fragile according to Angell's definition [40].

Recent experimental results obtained by Sokolov and coworkers [41] studying
polyisobutylene and polystyrene show the presence of a critical behavior only
above $T_{\text{MC}}$\ along with the failure of the predictions of the mode
coupling theory below the critical temperature. These authors show many
similarities between the results obtained on these polymeric glasses and the
ideas of the liquid-liquid transition in polymeric liquids formulated by Boyer
and coworkers [42,43]. It is worth noting that this liquid-liquid transition
would manifest itself through a discontinuity in the first derivative of the
specific heat (and not the specific heat itself as in the present case) at the
transition temperature. This makes Boyer's result very different from our
result. The idea introduced by Boyer has been strongly criticized and is still
the subject of discussion [44,45].

The second similarity has to do with our choice of the parameter $a$, so that
$T_{\text{MC}}$ lies above the Kauzmann temperature $T_{\text{K}}.$ This is
also what is expected in the mode-coupling theory in which the transition
occurs above the conventional glass transition temperature.

The third similarity appears when we allow free volume in our model in Eq.
9(b), as has been done recently [29]. It is found that the free volume in the
model for the case of infinite polymers \textit{vanishes} identically at
$T_{\text{MC}}$, and remains zero below it. Consequently, one expects the
viscosity to diverge at $T_{\text{MC}}$.

While the mode coupling theory describes the transition at $T_{\text{MC}}$\ as
dynamic in nature, our results show that the transition at $T_{\text{MC}}$ is
thermodynamic in nature. The sharp transition is due to the polymer being
infinitely long, and disappears as soon as polymers become \textit{finite} in
size [29]. However, for polymers of reasonable sizes, there will continue to
be a narrow crossover region between two phases (ML and SCL).

\subsection{Ground State and Kauzmann Temperature}

Below $T_{\text{MC}}$, the only two phases that are present are the metastable
liquid and the crystal. Above $T_{\text{MC}}$, the supercooled liquid, which
is the continuation of the equilibrium liquid above $T_{\text{M}}$, is more
stable than the metastable liquid and coexists with the crystal. It is worth
noting that the modified free energies $\widetilde{F}$ of both the metastable
liquid and the supercooled liquid cross over and becomes \textit{positive} at
some finite and non-zero temperature. Let us focus on the metastable liquid as
its behavior is easy to describe since its modified free energy $\widetilde
{F}$ \ remains \textit{independent} of$\ a$ and $b$. We first observe that the
2-cycle FP solution contains within its possible solutions the 1-cycle
solution. We also find that the free energy of all possible 2-cycle solutions
(including the 1-cycle solutions) at absolute zero are the \textit{same}:
$\widetilde{F}=0.$

Because of the exact nature of our calculation, this equality of the ML and CR
free energies at absolute zero is not brought about due to any approximation.
Because of this equality at absolute zero, we will now consider the modified
free energy $\widetilde{F}$ in the following. The CR free energy remains
negative at all temperatures and approaches zero at absolute zero. Thus,
CR\ has non-negative entropy. On the other hand, the ML free energy
$\widetilde{F}$, which is negative at higher temperatures, becomes positive at
some intermediate and non-zero temperature $T_{\text{M0}} $ and keeps
increasing, as the temperature is lowered, until the Kauzmann temperature
$T_{\text{K}}$ ($\approx0.48$ in our model) is reached. Below $T_{\text{K}}$,
the ML\ free energy $\widetilde{F}$ \ must necessarily decrease since it must
vanish at absolute zero. The maximum of $\widetilde{F} $ corresponds to the
vanishing of the entropy of the ML phase, below which the entropy must become
\textit{negative} [46]. Thus, the existence of the Kauzmann temperature is a
consequence of the fact that the ML\ free energy $\widetilde{F}$ , once it
crosses the zero at $T_{\text{M0}}$, must necessarily decrease at some lower
temperature so as to return to zero at absolute zero. The existence of the
maximum in $\widetilde{F}_{\text{ML}}$ as a function of \ the temperature is
the root of the rapid entropy drop noted by Kauzmann [5]. The maximum at a
positive $T_{\text{K}}$ is forced by thermodynamics since the larger specific
heat of ML makes $\widetilde{F}_{\text{ML}}$ cross over to positive values at
$T_{\text{M0}}$. If we had carried out our calculation in some approximation,
as is the case with the calculation of Gibbs and DiMarzio [8], we certainly
could not draw this remarkable conclusion.

\subsection{$\mathit{S}_{\text{CR}}\mathbf{>}\mathit{S}_{\text{SCL}}$ and
Entropy Crisis}

The crystalline phase has an entropy that is never negative. Hence, its
entropy is larger than the entropy of the metastable liquid in the temperature
interval $T_{\text{K}}<T<T_{\text{eq}}$, where $T_{\text{eq}}$\ is the
temperature at which the entropy of the two phases is the same (see Fig.
10(b)). This result contrasts the common belief [17] that the entropy of a
crystalline phase must always be lower than the entropy of the corresponding
liquid phase, even if the latter is an equilibrium phase. Our results clearly
show that there is no thermodynamic requirement for this belief to be true.
Indeed, real systems like He conform with this observation.

In order to sustain the common belief that the entropy of the liquid must
always be larger than that of the crystal, it was conjectured by Kauzmann that
the system must avoid the (Kauzmann) catastrophe caused as soon as the
requirement $S_{\text{CR}}\leq S_{\text{SCL}}$ is violated. The system is
supposed to avoid the catastrophe by undergoing either a spontaneous
crystallization, as proposed by Kauzmann in his original paper, or an ideal
glass transition [5,8,15,47].

The most important result of the present research regarding the Kauzmann
catastrophe is that since our calculations show that it is possible to observe
$S_{\text{CR}}>S_{\text{SCL}}$ in an explicit model calculation, the existence
of the catastrophe must be reinterpreted in terms of the entropy crisis
corresponding to having no realizable state\ (negative entropy) [46]: the
catastrophe happens at a temperature $T_{\text{K}}$\ where the entropy of the
metastable liquid goes to zero and not at the temperature $T_{\text{eq}}$ at
which the entropy of the liquid becomes equal to the entropy of the crystal.
The two temperatures coincide in the original Flory model since the entropy of
the crystal is identically equal to zero below $T_{\text{M}}$, see Fig. 1,
while in our exact calculation, the two temperatures are different because the
entropy of the crystal is zero only when the temperature goes to zero.

\subsection{Ideal Glass Transition}

The ML free energy has no physical relevance below $T_{\text{K}}$\ since it
corresponds to negative values of the entropy. Decreasing the temperature as
$T\rightarrow T_{\text{K}}^{+}$ the entropy falls very rapidly to zero (as
seen, for example, in Fig. 10 (b)). As the temperature is decreased, the
energy of the metastable liquid cannot increase because this would correspond
to a negative specific heat. At the same time, this energy cannot decrease
since there are no states with non-negative entropy (except the crystal phase,
but we do not consider this possibility here) available to the system. The
only conclusion that can be drawn from these observations is that for
$T<T_{\text{K}}$, the metastable liquid remains frozen in the state in which
it finds itself at $T_{\text{K}}$. \textit{This describes the ideal glass
transition}. In our analysis, ML does not undergo any changes in its state at
the glass transition, so that this transition cannot be a first-order
transition as recently proposed by Parisi and coworkers [26(b)].

It is interesting to notice that the energy of the metastable liquid increases
monotonically with $T$ and is very similar to the excitation profiles for
other systems like the mixed Lennard-Jones system [48] and the two states
model [49].

It is also interesting to notice that our model predicts, at the Kauzmann
temperature, an upward discontinuity for the specific heat. The specific heat
of the metastable liquid is decreasing with the temperature for $T>T_{\text{K}%
}$. This kind of behavior is in disagreement with many experimental results
but has been observed in computer simulations by Parisi and coworkers [26] and
experimentally (at least in some temperature interval immediately above the
glass transition) in glasses formed by low molecular weight materials of very
different nature like 1-butene [45] and the metallic system Au-Si [40].

\subsection{Flory Model}

The value $a=1$ corresponds to a borderline case. As we have shown in Fig. 4,
if $a$ is larger than 1 then the ground state is not crystalline anymore and
there are no parallel bonds at $T=0$. Correspondingly, we observe from our
results that as $a\rightarrow1$, $T_{\text{MC}}$\ moves to lower and lower
values and eventually goes to zero when $a$ goes to one. We also find that, if
we keep $b$ fixed, then as the value of $a$ increases, the values of
$T_{\text{M}}$\ and $T_{\text{MC}}$\ decrease and, as said above, for
$a\geq0.8$, the supercooled liquid shows its own Kauzmann temperature
corresponding to the maximum in its free energy. If $a$ becomes negative, the
critical temperature moves to higher temperatures. Unlike the case of $a>0$,
there is a temperature interval $(T_{\text{M}}<T<T_{\text{MC}})$ where the
metastable liquid becomes the stable phase EL$_{\text{ML}}$. The melting
transition that is observed is a continuous transition. This aspect is in
disagreement with experimental observations.

If $a=b=0$, as explained before, the model reduces to the Flory model of
polymer melting. In this case $T_{\text{M}}=T_{\text{MC}}=T_{\text{CRE}}$ and
the melting transition from the equilibrium liquid to the crystalline phase
becomes continuous in contrast with the original calculations of Flory. It is
important to notice that in this case there is no possibility to obtain a
supercooled liquid since the equilibrium liquid phase disappears continuously
into the crystal at $T_{\text{M}}$.

The solution found by Flory for the original model corresponds to a
tricritical solution: if we consider the melting transition present in the
system, this transition is first order for $a>0$ and continuous for $a<0$. In
the latter case, we have $T_{\text{M}}\equiv T_{\text{CRE}}$, as shown in Fig.
14.%
\begin{figure}
[tb]
\begin{center}
\includegraphics[width=4.5in]
{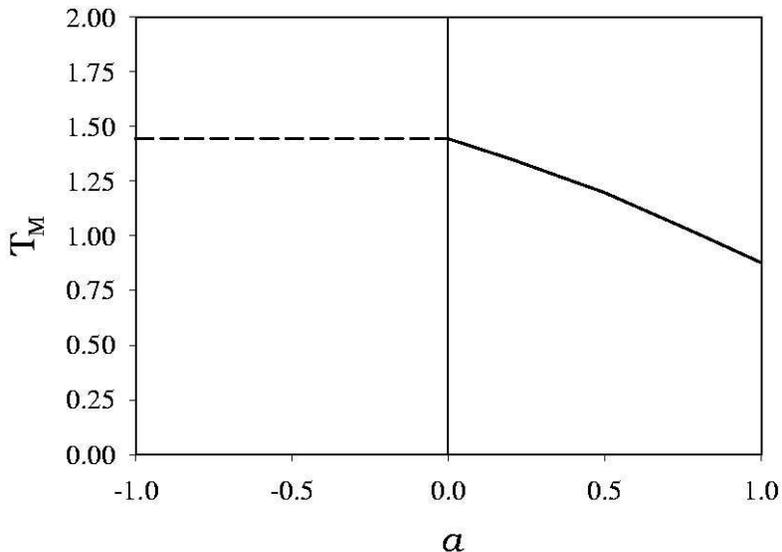}%
\caption{Dependence of the melting temperature on $a$. The first-order
transition line (continuous line ) and the second-order transition line
(dashed line) are shown.}%
\end{center}
\end{figure}

It seems reasonable to assume that in order to be able to describe the physics
of real systems, the value of the parameter $a$ must be chosen in the range
between 0 and 0.7 while the value of $b$ should be in the range $-0.5-0.5$.

\subsection{Comparison with A Real System}

By a proper choice of various parameters in our model, we can fit the
predictions of our theory with experiments. We discuss one such example below.
Setting $a=0.5$\ and $b=0$, for example, our model predicts $T_{\text{M}%
}/T_{\text{k}}\simeq1.20/0.48$. We can consider polyethylene (PE) and try to
describe its thermodynamic properties using our model. The experimentally
measured melting temperature of PE is $T_{\text{M}}($PE$)\simeq400$K. Then the
model predicts $T_{\text{K}}($PE$)\simeq160$K. This temperature is about 40K
below the experimentally determined glass transition. Since we expect the
experimental glass transition to occur above the ideal glass transition
because of experimental constraints, we can conclude that the prediction of
our model is, at least, reasonable.

\section{CONCLUSIONS}

We have considered an extension of the Flory model of melting by introducing
two additional interactions characterized by parameters $a$ and $b$. One
interaction is between a pair of parallel bonds and the second one is due to a
hairpin turn. The model is defined on a checkerboard version of the square
lattice, and has been solved \textit{exactly} on a Husimi cactus, which is a
recursive lattice [36]. It should be recalled [36] that calculations done on a
recursive lattice have been shown to be more reliable than conventional
mean-field calculations. The choice of the Husimi cactus is also important for
the inclusion of the Gujrati-Goldstein excitations that are responsible for
destroying the complete order in the crystal phase CR [10-12] in the Flory
model. The method of calculation is to look for the fix-point (FP) solution of
the recursion relations. We need to consider 1-cycle and 2-cycle FP solutions
to describe the disordered phase and the crystal phase, respectively. This has
required us to provide in this work the extension of the Gujrati trick [36] to
calculate the free energy of the 1-cycle FP solution to the 2-cycle solution.
The exact nature of the solution allows us to draw some important conclusions,
which would not have been drawn with the same force, had we obtained the
solution under ambiguous and/or uncontrollable approximations.

We have identified an equilibrium liquid phase EL at high temperatures $T\geq
T_{\text{M}}$. There is another liquid phase ML which exists at all
temperatures, but which never\ becomes an equilibrium phase for appropriate
choices of the parameters $a$ and $b$. Below the melting temperature
$T_{\text{M}}$, the crystal phase CR becomes the equilibrium phase. The
melting transition is usually a first order transition with a latent heat,
below which we can continue EL to give rise to the supercooled liquid phase
SCL. This phase terminates at a lower temperature $T_{\text{MC}}$, where it
meets ML\ continuously with no latent heat. The transition at $T_{\text{MC}}$
is a continuous liquid-liquid transition.

Both SCL and ML represent metastable phases in the system. For a metastable
state to exist in Nature, its entropy must be non-negative. A negative entropy
($S<0)$ in metastable states (SCL\ and/or ML) implies that such states
\textit{cannot} exist. We have argued that the ML\ free energy must have a
maximum at $T_{\text{K}}$ as a function of the temperature $T$; see Fig. 8.
Thus, ML has non-negative entropy above $T_{\text{K}}$, but gives rise to
negative entropy below $T_{\text{K}}$. We have called the appearance of $S<0$
the entropy crisis , which we have used instead of the Kauzmann paradox
($S_{\text{ML}}<S_{\text{CR}}$) as the driving force behind the ideal glass
transition at $T_{\text{K}}$. The ideal glass transition occurs in the
metastable liquid ML, and not in SCL, which is contrary to the conventional
wisdom. The portion of ML below $T_{\text{K}}$ must be replaced by the ideal
glass (see dotted thin horizontal portion below $T_{\text{K}}$ in Fig. 8),
which is completely inactive in that its entropy and specific heat are both
zero. The rapid drop in the entropy near the Kauzmann temperature
$T_{\text{K}}$ is a direct consequence of the existence of the maximum in the
ML free energy. The liquid-liquid transition at $T_{\text{MC}} $ between the
two metastable phases ML and SCL has been shown to share many similarities
with the critical mode-coupling transition, even though the latter is known to
be driven by dynamics in simple fluids. It should be remarked that nothing is
known about this dynamic transition in the \textit{semiflexible Hamilton walk}
limit of infinite polymers.

The Flory model is shown to give rise to a continuous melting. Indeed, the
melting point in the Flory model turns out to be a tricritical point in our
calculation. It would be interesting to see if this conclusion can be
sustained in other computational scheme.

The natural extension of this model in Eq. (9b) involves the analysis of the
effects of compressibility (taking into account voids as another species on
the lattice) and of finite chain size, both in the polydisperse and in the
monodisperse case. This is reported elsewhere [29,30].

We finally discuss an interesting aspect of the 2-cycle FP solution \ for
EL/SCL observed by Semerianov [50]. The values of $i_{\mathrm{a}}$,
$i_{\mathrm{b}}$, $o_{\mathrm{a}}$, and $o_{\mathrm{b}}$ depend on the choice
of the initial guesses used in the FP solution of the recursion relations.
Thus, there are many different 2-cycle solutions for EL/SCL that differ in the
values of $i_{\mathrm{a}}$, $i_{\mathrm{b}}$, $o_{\mathrm{a}}$, and
$o_{\mathrm{b}}.$What we find is that the product $io$ remains the same on
both sublattices for all initial guesses, and that they all give the
\textit{same }free energy and densities. We hope that the observation will
provide some useful information about the free energy landscape picture, but
this requires further investigation. We hope to report on this in near future.

\begin{center}
\textbf{ACKNOWLEDGEMENTS}
\end{center}

We would like to thank Sagar Rane and Fedor Semerianov for fruitful discussions.

\pagebreak 

\begin{center}
\textbf{APPENDIX}
\end{center}

We index the cactus levels in a slightly different manner here for simplicity.
If the base site of a square is indexed $m$, then the intermediate and the
peak sites are all indexed $m+1$. The square is still called the $m$th square.
The origin is indexed, as before, $m=0$. We evaluate $N_{(m)}$, the number of
sites at the $m$th generation of a tree rooted at the origin ($m=0$). The
rooted tree is only half of a complete tree. It is evidently given by:
\[
N_{(m)}=3^{m},
\]
so that the total number of sites belonging to the first $m$ generations (and
excluding the origin) of a rooted tree is:
\[
N_{(m)}^{rooted}=\overset{m}{\underset{k=1}{\sum}}N_{(k)}=\frac{3^{m+1}-3}%
{2}.
\]
The total number of sites belonging to the first $m$ generations of the
complete tree is then equal to twice the total number of sites belonging to
the first $m$ generations of a rooted tree plus one since we have to consider
the origin too:
\[
\widetilde{N}_{m}=2N_{(m)}^{rooted}+1=3^{m+1}-2.
\]

Let us consider now the number $S_{(m)}$ of the$\ m$th generation squares
($m\geq0$) of the rooted tree. Clearly we have:
\[
S_{(m)}=3^{m}.
\]
The total number of squares belonging to the first $m$ generations of a rooted
tree (so that the maximum generation of the squares is $m-1$) is:
\[
S_{(m)}^{rooted}=\overset{m-1}{\underset{k=0}{\sum}}S_{(k)}=\frac{3^{m}-1}%
{2}.
\]
The total number of squares belonging to the first $m$ generations of the
complete tree is then just twice the total number of squares belonging to the
first $m$ generations of a rooted tree:
\[
\widetilde{S}_{m}=2S_{(m)}^{rooted}=3^{m}-1.
\]

In order to make the cactus homogeneous, we must consider it to be part of a
larger cactus. Consequently, both $\widetilde{S}_{m}$ and $\widetilde{N}_{m}$
must be modified in order to take into account the presence of dangling bonds
at surface sites: we consider to add half a square to each surface site.
Recalling that each square contains four half-sites (a site is shared by two
squares), we conclude that each square contributes two sites to the number of
sites. Hence, each half-square contributes one site to the site count. Thus,
we modify $\widetilde{N}_{m}$ by adding 1 site for each surface square. This
gives for the total number of sites
\[
N_{m}=\widetilde{N}_{m}+3^{m}=4\times3^{m}-2.
\]

Modifying $\widetilde{S}_{m-1}$ by the half-squares at the surface sites gives
for the total number of squares:
\[
S_{m}=\widetilde{S}_{m-1}+\frac{1}{2}\times2\times N_{(m)}=2\times3^{m}-1.
\]

Now, if we consider the thermodynamic limit in which $m\rightarrow\infty$ we
have:
\[
\frac{S_{m}}{N_{m}}\rightarrow\frac{1}{2}.
\]
which is consistent with our earlier homogeneous hypothesis $N_{\text{S}}=N$
\ relating the total number of squares $N_{\text{S}}$ and the total number of
sites $N$.

Let us finally consider the total number $\widetilde{B}_{m}$ of bonds in the
first $m$ generations ($m>0$) \ of the tree. Each square contains four bonds,
and there are $\widetilde{S}_{m}$ squares in this tree. Thus
\[
\widetilde{B}_{m}=4(3^{m}-1).
\]
The modification of the lattice introduced above implies that we must now add
half-square at each of the surface sites of the $m$th generations ($m>0$)
tree. Each half-square contributes 2 bonds. Hence
\[
B_{m}=\widetilde{B}_{m}+4\times3^{m}=8\times3^{m}-2.
\]
If we consider the thermodynamic limit in which $m\rightarrow\infty$ we have:
\[
\frac{B_{m}}{S_{m}}\rightarrow4,
\]
as expected.

\pagebreak


\begin{thebibliography}{99}
\bibitem{Ref 32}In classical statistical mechanics, which is what is
conventionally used to analyze metastability, the contribution
$S_{\text{trans}}(T)$ to the total entropy $S(T)$ from the translational
degrees of freedom is the same for various possible phases like SCL or CR that
can exist at a given temperature, and is a function only of the temperature
$T$. This is because the kinetic energy is decoupled from the interaction
energy in classical mechanics, making the translational degrees of freedom
decouple from the rest [31(a)]. Thus, the translational degrees of freedom do
not have to be considered in the investigation. The difference $S_{\text{conf}%
}(T)\equiv S(T)-S_{\text{trans}}(T)$ can be equally well used to characterize
various phases of the system; it is known as the \textit{configurational
entropy} of the system and determines the configurational partition function.
In lattice models like the one considered here, we do not have the
translational degrees of freedom, and the translational entropy is zero.
Hence, there is no distinction between the total entropy $S(T)$\ and the
configurational entropy $S_{\text{conf}}(T)$\ in lattice models. For this
reason, we will not use the subscript ''conf'' to denote the configurational
entropy in this work.
\end{thebibliography}
\end{document}